\documentclass[12pt,preprint,round,colon,apj]{emulateapj}
\begin{document}

\shortauthors{Metchev \& Hillenbrand}
\shorttitle{Palomar AO Survey of Young Stars}

\title{Initial Results from the Palomar Adaptive Optics Survey of Young 
Solar-Type Stars:  a Brown Dwarf and Three Stellar Companions}

\author{Stanimir A.\ Metchev and Lynne A.\ Hillenbrand}

\affil{California Institute of Technology}
\affil{Division of Physics, Mathematics \& Astronomy, MC 105-24, Pasadena, 
California 91125}
\email{metchev, lah@astro.caltech.edu} 

\slugcomment{Accepted for publication in ApJ, Aug 27, 2004}

\begin{abstract}

We present first results from the Palomar Adaptive Optics 
Survey of Young Stars conducted at the Hale 5~m telescope.
Through direct imaging we have discovered a brown dwarf and two 
low-mass stellar companions to the young solar-type stars 
HD~49197, HD~129333 (EK~Dra), and V522~Per, and confirmed a 
previously suspected companion to RX~J0329.1+0118 
\citep{sterzik_etal97}, at respective separations of 0.95$\arcsec$
(43~AU), 0.74$\arcsec$ (25~AU), 2.09$\arcsec$ (400~AU), and 3.78$\arcsec$
(380~AU).  Physical association of each binary system is 
established through common proper 
motion and/or low-resolution infrared spectroscopy.
Based on the companion spectral types, we
estimate their masses at 0.06, 0.20, 0.13, and 0.20~$M_\sun$,
respectively.  From analysis of our imaging data combined with archival 
radial velocity data, we find that the spatially resolved companion to 
HD~129333 is potentially identical to the previously identified 
spectroscopic companion to this star 
\citep{duquennoy_mayor91}.  However, a discrepancy with the absolute
magnitude suggests that the two companions could also be
distinct, with the resolved one being the outermost component of a
triple system.  The brown dwarf
HD~49197B is a new member of a growing list of directly imaged sub-stellar
companions at 10--1000~AU separations from main sequence stars, 
indicating that such brown dwarfs
may be more common than initially speculated.

\end{abstract}


\section{INTRODUCTION \label{sec_intro}}

High-contrast imaging searches for low-mass companions to nearby and/or young
stars have increased dramatically in number since the initial
discovery of a brown dwarf companion to a main sequence star
(Gl~229) through direct imaging \citep{nakajima_etal95}.  One
particularly powerful technique is adaptive optics (AO), which provides
the high angular resolution ($\lesssim0.1\arcsec$) achievable at the
diffraction limit of large ground-based telescopes. The
widening niche of high-contrast imaging opened by recent developments in AO
technology implies that not only brown dwarfs but exo-solar planets may be
within the realm of direct imaging.
 Nowadays nearly every ground-based telescope equipped with an AO system
hosts an imaging companion-search project.  The sudden
explosion in interest in this topic has been fueled by the success of the
radial velocity (r.v.) method in detecting solar system analogs
\citep[e.g.,][]{marcy_butler98}.
Through longer time-lines of observation and higher
precision, the sensitivity of r.v.\ surveys has now extended outwards to
include planets at semi-major axes $\gtrsim$3~AU
\citep{carter_etal03,naef_etal04}, i.e., near the Jovian region in the
Solar System.  While the sensitivity of direct imaging
to ``planetary-mass'' \citep[1--13 Jupiter masses
($M_J$);][]{burrows_etal97} objects at
such separations from Sun-like stars is still extremely limited due to 
contrast requirements,
several higher-mass brown dwarf companions have been discovered at wider
separations \citep[see compilation in][]{reid_etal01}, some at projected
distances as small as 14--19~AU \citep{els_etal01,liu_etal02}.
High-resolution spectroscopic monitoring and direct imaging are thus
complementary in
searching for sub-stellar companions to stars. Future development of
both methods promises to narrow, and eventually close the sensitivity gap
between them.

Young stars are the most suitable targets for direct imaging of
sub-stellar companions.  At ages of 10--100 
million years (Myr) the expected brightness ratio in the near IR between a 
10~$M_{\rm Jup}$ object and a
solar-type star is $10^{-3}$--$10^{-5}$ \citep{burrows_etal97,baraffe_etal03}.
Modern AO systems can routinely achieve comparable dynamic range at 
1$\arcsec$ separations from bright stars.  Hence, for young stars within
40~pc of the Sun we can probe for massive planets at separations
comparable to the giant planet region in our own Solar System.
However, few young stars are known at such small heliocentric distances.
These are constrained to several tens of members of young moving 
groups: TW~Hya \citep{rucinski_krautter83,delareza_etal89}, Tucana/Horologium 
\citep{zuckerman_webb00,zuckerman_etal01a}, and $\beta$~Pic 
\citep*{zuckerman_etal01b}, and have already been targeted with
sensitive space-based and ground-based AO surveys, which have uncovered
3--4 brown dwarf companions \citep{lowrance01, lowrance_etal03,
neuhauser_guenther04}, but no planetary-mass ones.

Because contrast and projected separation are the limiting factors in
detectability of sub-stellar companions, brown dwarfs, 
being more luminous than planets, are detectable at 
greater heliocentric distances and at smaller angular separations 
from their host stars.
At the same physical separation from the primary (e.g., 50--100~AU),
brown dwarf companions should be detectable around older (several 
gigayears [Gyr]) and/or more distant ($\lesssim$200~pc) stars compared
to planets, allowing a larger sample of targets.  With regard to this,
we have commenced a survey of young
($<$400~Myr) solar-type (F5--K5) stars within 160~pc using the AO system on 
the Palomar 5-m telescope.  Our survey sample is largely a subset of
the sample targeted by the Formation and Evolution of Planetary Systems
(FEPS) {\it Spitzer} Legacy Team \citep{meyer_etal04}.
Although faint primary stars, such as M dwarfs
or white dwarfs, offer more favorable
contrast for imaging sub-stellar companions, we have chosen
to limit our sample to solar analogs because of interest in determining
the multiplicity statistics of sub-stellar objects around other suns.
Furthermore, several recent large 
surveys have already explored the multiplicity of nearby ($\lesssim$50~pc)
cool stars \citep{close_etal03,carson_etal03, mccarthy_zuckerman04}, or
white dwarfs \citep*{zuckerman_becklin92, green_etal00},
while a large sample
of F--G stars has not been studied, because of comparatively small 
numbers in the immediate solar neighborhood.

Preliminary results from our survey were reported in
\citet*{metchev_etal02}.  Here we present the strategy of the survey, and
the discovery and confirmation
of resolved low-mass companions to HD~49197, HD~129333 (EK~Dra), V522~Per, 
and RX~J0329.1+0118.  We shall refer to these throughout the paper as
HD~49197B, HD~129333B, V522~PerB, and RX~J0329.1+0118B.  For convenience
of notation, a second candidate companion to HD~49197 
found to be an unrelated background star will be denoted as
HD~49197``C''.  The full sample and further results from the survey will 
be discussed in a later paper.

\section{OBSERVING STRATEGY}

The observations described in this Section are representative of our
general survey observing strategy.  Table~\ref{tab_observations} details
the imaging and spectroscopic observations specifically for the four 
objects presented
here.  The properties of the observed primaries are given in 
Table~\ref{tab_properties}.

\subsection{Imaging}

\subsubsection{First-Epoch Imaging and Survey Sample Subdivision
\label{sec_1ep}}

First epoch observations are obtained with the
Palomar AO system \citep[PALAO;][]{troy_etal00} in residence at the
Cassegrain focus of the Palomar 5-m telescope.  Since the summer of 2003
the wavefront sensor runs at frame rates up to 2~kHz, and the system
routinely produces
diffraction-limited images (0.09$\arcsec$ at $K_S$) with 
Strehl ratios in the 30--50\% range at 2$\micron$ on 
$V<12$ guide stars, and up to 75\% on
$V<7$~mag guide stars.  PALAO employs the Palomar High Angular
Resolution Observer \citep[PHARO;][]{hayward_etal01}, a
1024$\times$1024~pix HgCdTe HAWAII detector with imaging (25~mas/pix and
40~mas/pix plate scale) and spectroscopic ($R$=500--2000) capabilities in
the near IR.  A set of coronagraphic spots, Lyot masks and neutral
density (ND) filters are available to achieve the desired dynamic
range.  

Our program entails $K_S$-band (2.15$\micron$) imaging in the 25~mas/pix 
mode (25$\arcsec\times25\arcsec$ field of view) both 
with and without a 0.97$\arcsec$-diameter coronagraphic stop.
For high dynamic range, long (1~min) coronagraphic images are taken to 
identify fainter 
(potentially sub-stellar) companions at separations $>$0.5$\arcsec$.
Twenty-four such exposures are taken, for a total of 24~min
integration per target, with 6~min spent at 
each of 4 different orthogonal detector orientations (obtained by
rotating the Cassegrain ring of the telescope).  
For every 6 min of on-target imaging (i.e., at each detector
orientation), separate 1-min coronagraphic 
exposures are taken at five dithered sky positions
32--60$\arcsec$ from the star.
For high angular resolution (but with lower dynamic range), short 
(1.4--9.8~sec) non-coronagraphic exposures are taken
to look for close companions of modest flux ratio, and to establish 
relative photometric
calibration.  The images are taken in a 5-point dither
pattern at the vertices and center of a box 6$\arcsec$ on a side.  
A 1\%-transmission ND filter is used if necessary to avoid 
saturation\footnote{The ND~1\% filter was calibrated photometrically 
through repeated
(17--20 per band) observations of 3 program stars with and without the 
filter, and its extinction was measured at 4.753$\pm$0.039~mag at $J$,
4.424$\pm$0.033~mag at $H$, and 4.197$\pm$0.024~mag at $K_S$.}.
On occasion, a narrow-band (1\%) Brackett-$\gamma$ (2.17$\micron$)
filter is used for higher throughput, instead of the ND~1\% filter.

To avoid detector saturation and/or decreased sensitivity over a 
substantial fraction of
the image area, stars with bright ($\Delta K_S<4$) projected
companions in the PHARO field of view (FOV) were not observed with deep
coronagraphic exposures.  However, binaries with separation $\leq0.5\arcsec$
were included, as both components of the binary could then be
occulted by the coronagraph.  This naturally splits our survey sample in two
groups: the ``deep'' subsample, consisting of essentially single stars
and close binaries, and the ``shallow'' subsample encompassing the remaining
stars.  Membership to one of the two subsamples was assigned at the telescope,
when their multiplicity and approximate flux ratio was revealed during the
short exposures.  The shallow subsample was further expanded to include 
stars out to 200~pc and/or 
older than 400~Myr to cover the entire FEPS sample accessible from the
Northern hemisphere.  

Short dithered exposures were taken of all stars, while long
coronagraphic exposures were taken only of stars in the deep subsample
at $K_S$ band.
In addition, short $J$- (1.22$\micron$) and $H$-band (1.65$\micron$) 
exposures were taken of all candidate binaries (all stars in the shallow
survey, and the $<0.5\arcsec$ systems in the deep survey)
to allow approximate photospheric characterization of the components.  

In accordance with the above distinction, HD~49197 was observed for a
total of 24~min with the coronagraph as a part of the deep survey, 
while HD~129333 and
RX~J0329.1+0118 (with bright candidate companions), and V522~Per 
\citep[$\alpha$~Per member, 190~pc from the Sun;][]{vanleeuwen99} were
observed only with short exposures.  Conditions were
photometric during the first epoch observations of HD~49197, 
V522~Per, and RX~J0329.1+0118,
and unstable during those of HD~129333.  

\subsubsection{Follow-Up Imaging \label{sec_followup}}

We obtain second-epoch imaging
observations of all candidate companions to check for common proper 
motion with their corresponding stars.  Such were taken
for HD~129333 with PALAO/PHARO, and for HD~49197, V522~Per, and RX~J0329.1+0118
with NIRC2 (Matthews et al., in prep.) and the Keck~II AO system 
\citep[diffraction limit
0.05$\arcsec$ at $K_S$;][]{wizinowich_etal00}.  
Conditions were not photometric during follow-up, and only the best
images (Strehl ratio $S\gtrsim40\%$) were selected for astrometry.  
HD~129333 was followed up in the narrow-band Brackett~$\gamma$ 
filter, which allowed higher throughput than the ND~1\% filter
in the shortest (1.4~sec) PHARO exposures.  Given the unstable
atmospheric conditions during the second-epoch imaging of HD~129333,
this allowed us to take high signal-to-noise ($S/N$) exposures on 
time-scales that would most finely
sample the variations in the seeing, and to select only the ones with the
best imaging quality.

Keck follow-up is done at $JHK_S$, or at $K_S L^\prime$ if the candidate
companion is expected to be bright enough to be seen at $L^\prime$ 
\citep[given $0.7<K_S-L^\prime<2.5$ for L and T
dwarfs;][]{leggett_etal02}.  Sequences of short (up to 20~sec from
multiple co-adds) dithered non-coronagraphic, and long (1~min)
target-sky-target exposures are taken with a 1$\arcsec$- or 
2$\arcsec$-diameter coronagraph
in the same manner as with PHARO, though without detector rotations.  
The candidate companions
are exposed until a $S/N$ ratio comparable to that in the first-epoch 
PHARO observation is achieved (for
similar positional accuracy), up to 6~min per filter in $J$, $H$, 
and $K_S$.  The total integration time at $L^\prime$ is up to 10.5~min,
which allows the detection of $L^\prime\lesssim15.0$ objects.
We mainly use the 40~mas/pix 
(wide) NIRC2 camera (41$\arcsec$ FOV), which severely under-samples the 
Keck AO point-spread function (PSF), but is known
to suffer from less distortion than the 20~mas/pix 
(medium) camera over the same field \citep*{thompson_etal01}.  Although
we also have the option of using the 10~mas/pix (narrow) camera 
(10$\arcsec$ FOV) in NIRC2, it does not allow follow-up of distant
candidate companions, and we avoid using it for consistency with the
other NIRC2 observations.

\subsubsection{Imaging Data Reduction \label{sec_psf}}

All imaging data are reduced in a standard fashion for near IR
observations.  Flat fields are
constructed either from images of the twilight sky (for the Palomar
data), or from images of the lamp-illuminated dome interior (for
the Keck data).  A bad pixel mask is created from the individual flats,
based on the response of each pixel to varying flux levels.
Pixels whose gain deviates by more than 5 sigma 
from the mode gain of the array are flagged as bad.
Sky frames for the dithered, non-coronagraphic exposures 
are obtained by median-combining four of the five
exposures in the dither pattern (excluding the central pointing), and
rejecting the highest pixel value in the stack.  
The coronagraphic-mode sky frames are median-combined using an average
sigma clipping algorithm to remove pixels deviant by more than 5 sigma.
The sky-subtracted images of each target are divided by the
flat field, then registered, and median-combined to create a final high
signal-to-noise ($S/N$) image (Figures~\ref{fig_hd49197} and
\ref{fig_stellar}).  However, photometric and astrometric 
measurements are performed on the individual reduced images.  

No PSF stars are observed at either Palomar or Keck.  With PHARO at
Palomar, median-combined images from all 4 detector orientations
can be used to reproduce an approximate PSF.
This approach was chosen to emulate the observation of separate PSF
stars of identical brightness and color, while optimizing the time 
spent on science targets.  
However, we have found that a simple 180$\degr$ rotation and
subtraction technique works equally well, and we use that on both
the Palomar and Keck data.  
While neither approach eliminates telescopic speckle noise (as could be
the case if actual PSF stars were observed), both significantly
reduce point-symmetric structure in the PSF.

\subsection{Astrometric Calibration}

The exact pixel scale of the
25~mas PHARO camera was determined using known binary stars from the
Sixth Orbit Catalog \citep*{hartkopf_etal01,hartkopf_mason03a}:
WDS~09006+4147 \citep*[grade 1;][]{hartkopf_etal96}, WDS~16147+3352 
\citep[grade~4;][]{scardia79}, WDS~18055+0230
\citep[grade~1;][]{pourbaix00}, and WDS~20467+1607
\citep[grade~4;][]{hale94}.  These ``calibration binaries'' are observed
throughout our campaign at Palomar at all four detector orientations.
The combination of grade~1 (accurately
determined, short-period) orbits and grade~4 (less accurately known,
longer-period) orbits was selected from the list of astrometric calibrators
recommended by \citet{hartkopf_mason03b}.  Despite the lower quality of
the solution for binaries with grade~4 orbits,
their periods are generally much
longer (889 and 3249 years for WDS~16147+3352 and
WDS~20467+1607, vs.\ 21.78 and 88.38 years for WDS~09006+4147 and
WDS~18055+0230, respectively), so their 
motions are predicted with sufficient accuracy for many years into the
future.  The mean pixel scale of PHARO was measured to be 25.22~mas/pix
with a 1 sigma scatter of
0.11~mas/pix among measurements of the individual binaries at different
detector orientations.	This measurement is
consistent with, though less precise than our previous determination
\citep*[$25.168\pm0.034$~mas/pix;][]{metchev_etal03}, which was obtained
from only one calibration binary at a single Cassegrain ring
orientation.  The larger scatter of our more recent measurement is
indicative of the systematics involved in choosing different
calibration binaries, and in observing at more than one detector
orientation.

The plate scale of the wide NIRC2 camera was calibrated using the binary
WDS~15360+3948 \citep[grade~1;][]{soderhjelm99}.  The obtained value of
39.82$\pm$0.25~mas/pix is consistent with the pre-ship measurement of
39.7$\pm$0.5~max/pix\footnote{See 
http://www2.keck.hawaii.edu/inst/nirc2/genspecs.html.}.  Because only
one binary was used for calibrating NIRC2, a 0.63\% error
term corresponding to the uncertainty in the semi-major axis of the
binary has been added in quadrature to the uncertainty of our measurement.

\subsection{Spectroscopy \label{sec_obs_spec}}

\subsubsection{Observations}

Spectroscopy of faint targets in the halos of bright objects is
more challenging than spectroscopy of isolated targets.  Care needs to be
taken to achieve optimum suppression of the scattered light of the
bright stars either by 
using a coronagraph (in a manner similar to coronagraphic imaging), or by
aligning the slit so as to minimize light admitted from the halo.
Since neither PHARO nor NIRC2 allow coronagraphic spectroscopy, when
taking spectra of faint companions we orient the slit as 
close as possible to 90$\degr$ from the primary-companion axis.
Spectra of brighter companions, for which the signal from the halo
is negligible compared to that from the target, are obtained by
aligning the binary along the slit.  Given that we often use the 
F5--K5~IV--V primaries in our sample
as telluric standards, such alignment improves our observing
efficiency.  Sky spectra are obtained simultaneously with the target
spectra, by dithering the targets along the slit.

Promising candidate companions are observed at medium resolution
($R=1000-3000$) at $K$ and
(AO correction permitting) at $J$.  In PHARO we
use the corresponding grism and filter combination ($K$ or $J$)
and the 40~mas/pix camera, while in NIRC2 we use the $K$ or $J$ filter
with the ``medres'' grism, and the wide (40~mas/pix) camera.  For
$K$-band spectroscopy in first order with PHARO and in fifth order
with NIRC2 we achieve complete coverage of the 2.2$\micron$ atmospheric 
window.  For $J$-band spectroscopy in first order with PHARO the 
coverage is limited by the size of the detector to a 0.16$\micron$ 
bandwidth.  $J$-band spectroscopy with NIRC2 can cover the entire 
1.2$\micron$ window, but the spectrum is split between the fifth and 
the sixth dispersion 
orders.  To allow simultaneous data acquisition from both orders,
we fit both onto the wide camera by
clipping the fifth (longer-wavelength) order shortward of 1.22$\micron$ 
and the sixth (shorter-wavelength) order
longward of 1.28$\micron$.  The combined $J$-band spectrum has complete
coverage between 1.16--1.35$\micron$.

We took $J$ and $K$-band spectra with PHARO of 
HD~129333A/B, $J$ and $K$ spectra with NIRC2
of V522~PerA/B, and $K$-band NIRC2 spectra of
HD~49197A/B and RX~J0329.1+0118A/B.  For the observations of HD~129333 
with PHARO, and of RX~J0329.1+0118 with NIRC2, the
binary was aligned along the slit, whereas spectra of the other two 
binaries (with fainter companions) were obtained by placing
each component in the slit individually.  
A 0.26$\arcsec$ (6.5~pix) slit was used 
in PHARO, and a 0.08$\arcsec$ (2~pix) slit was used in NIRC2, resulting in 
spectral resolutions of 1200 at $J$ and 1000 at $K$
with PHARO, and 2400 at fifth-order $J$ (1.22--1.35$\micron$), 2900 at 
sixth-order $J$ (1.15--1.28$\micron$), and 2700 at $K$ with NIRC2.

Flat fields with the dispersive grisms in place were not
obtained for any of our spectroscopic observations.  Instead, 
the raw spectra were divided by an imaging flat field, constructed
in the same manner as for the imaging observations.
The flat-fielded spectra were corrected for bad 
pixels, using the same bad pixel mask as in the imaging case.
Strong positive deviations due to cosmic ray hits were then eliminated
using the L.A.~Cosmic Laplacian filter algorithm \citep{vandokkum01}.
Fringing was at a noticeable level ($\approx$15\%) only in spectra
obtained with PHARO (of HD~129333), and was reduced to below 5\% by 
dividing the target
spectrum by that of the telluric standard, taken at a nearby position on
the detector.

\subsubsection{Spectroscopic Data Reduction}

Extraction of the spectra is performed using {\sc iraf/apextract}
tasks.  To reduce contamination from the halo of the primary in the
companion spectra, the local background (arising mostly from the halo)
is fit by a first-order polynomial along the direction perpendicular to
the dispersion axis and subtracted during extraction.
In addition, the aperture width for the companion spectrum 
is conservatively set to the full width at half maximum of the profile so as
to include only pixels with maximum signal-to-noise.
Pixels within the aperture are summed along detector columns (which
are nearly perpendicular to the dispersion axis), and the resulting
``compressed'' spectrum is traced along the dispersion axis by fitting a high 
order (8 to 15) Legendre polynomial.  The tracing
step is approximately equal to the slit width, where at each step the data
point is obtained as the sum of 3--10 adjacent pixels (1.5--3 slit
widths; the spectra of fainter objects being more heavily averaged) in the 
compressed spectrum.
This procedure is found to produce consistent results for extractions
of multiple dithered spectra of the same companion, indicating that 
scattered light contamination from the primary has been reduced to a 
small level.  Nevertheless, in some cases of very faint close-in 
companions, the continuum shape is still found to vary noticeably
among the individual extractions.  We therefore avoid classifying the
spectra of the companions based on their continuum shapes, but
rely on the relative strengths of narrow absorption features (discernible
given our resolution) instead (Section~\ref{sec_spectroscopy}).

A dispersion solution for each spectrum is obtained from night sky lines in
non-sky-subtracted images,
using the task {\sc identify} and OH emission-line lists available in 
{\sc iraf}.  For the spectra of primaries observed separately (e.g., 
HD~49197 and V522~Per), whose shallower (10--120~sec) exposures do not contain
telluric emission lines at a high enough $S/N$ to allow the fitting of a
dispersion solution, such is derived after registration with 
deeper (10--15~min long) companion spectra taken
immediately after those of the primaries.  The tasks {\sc fxcor} and 
{\sc specshift} are 
used to cross-correlate and align the individual wavelength-calibrated
spectra for a given object.
The individual primary and companion spectra 
are then median-combined using {\sc scombine}.  

Since the
primaries (when earlier than K0) double as telluric standards, their 
spectra are first corrected for photospheric absorption from 
atomic hydrogen (H~P$\beta$ at 1.282$\micron$
and H~Br$\gamma$ at 2.166$\micron$) by hand-interpolating over the
absorption with the task {\sc splot}.  Weaker absorption features due
to \ion{Na}{1}, \ion{Ca}{1}, \ion{Si}{1}, \ion{Al}{1}, and \ion{Fe}{1}, 
although
present in the spectra of our telluric standards, are left uncorrected 
because of blending (at our resolution) with various OH absorption 
lines in the Earth's atmosphere.  After thus correcting the telluric spectra,
the spectra of the companions are divided
by those of the telluric standards.  Finally, the 
spectra of the companions are multiplied by black bodies
with temperatures corresponding to the spectral types of the telluric
standards \citep[based on effective temperature vs.\ spectral type data 
from][]{cox00}, and boxcar-smoothed by the width of the slit.  

The above reduction procedure applies in exact form for NIRC2 spectra,
which do not suffer from fringing.
For PHARO data, which are noticeably affected by fringing, we
divide the individual target spectra by those of
the telluric standard before median-combining and wavelength calibration,
since fringing depends on detector position, not on the dispersive element.

Reduced spectra for the objects reported here are presented in
Figure~\ref{fig_Kspectra} and \ref{fig_Jspectra}.  The mismatch in the
continuum slopes between the fifth- and sixth-order the $J$-band spectra 
of the companion to V522~Per may be due to our use of imaging, instead 
of spectroscopic flats.

\section{ANALYSIS}

\subsection{Photometry \label{sec_anal_photometry}}

Broad-band near-IR photometry of the companions is presented
in Table~\ref{tab_photometry}.  The measured quantity in each case is
the relative flux ($\Delta J$, $\Delta H$, $\Delta K_S$, $\Delta
L^\prime$) of the companion with respect to that of the primary.  
When the companions were visible without the coronagraph (in all cases
except for the close-in companion to HD~49197), the fluxes of both
components were measured directly from the short-exposure non-coronagraphic 
images.  Flux ratios for the HD~129333, V522~Per, RX~J0329.1+0118, and
HD~49197A/``C''
systems were obtained from the Palomar images in apertures of radii
of 0.2$\arcsec$ ($2\lambda/D$ at
$K_S$). For the photometry of HD~129333B we subtracted the halo of
the nearby (0.74$\arcsec$) primary (as detailed in
Section~\ref{sec_psf}) to minimize its contribution to 
the flux of the secondary.  
In the case of the HD~49197A/B system, the 0.95$\arcsec$ companion is 
not seen in
PHARO images taken without the coronagraph, and only coronagraphic
exposures were obtained with NIRC2.  The magnitude of the
companion was obtained with
respect to the residual flux of the primary seen through the 1$\arcsec$
NIRC2 coronagraphic spot.  The flux measurements were performed on the
PSF-subtracted NIRC2 images in a 0.16$\arcsec$-radius (3.2$\lambda/D$
at $K_S$) aperture.  The transmission
of the spot was established from photometry of another program star
in images taken with and without the 2$\arcsec$ coronagraph: 
its extinction was measured at 9.27$\pm$0.07~mag at $J$, 
7.84$\pm$0.03~mag at $H$, and 7.19$\pm$0.03~mag at $K_S$.
The companion was seen in all 6
one-minute coronagraphic exposures at $K_S$, but because of varying 
photometric conditions and proximity of diffraction spikes, it was
detected in only 3 out of 6 PSF-subtracted exposures at $H$, and 1 
out of 6 at $J$.

Sky values for each of the observed objects were determined as the
centroid of the flux distribution in a
0.1$\arcsec$-wide annulus with an inner radius larger by $\lambda/D$
than the distance at which the radial profile of the object fell below
the level of the local background.  For the primaries observed with
PHARO
this inner radius was 2.0$\arcsec$ (20$\lambda/D$ at $K_S$), while for
HD~49197A, the flux of which was measured through the NIRC2 coronagraph,
the inner radius was 0.2$\arcsec$ (5$\lambda/D$).  For the
companions, the inner radius of the sky annulus varied from
2.25--4 times the $K_S$ diffraction limit, depending on whether the
local background was strongly
influenced by the halo of the primary (as near HD~49197 and HD~129333),
or not.	 Experiments with varying sky annulus
sizes in the two more detached systems (V522~Per and RX~J0329.1+0118)
showed that the relative photometry between two objects in the same
image is preserved to within 0.08~mag for
annuli ranging between 2.25--20$\lambda/D$ in inner radius.

The $J-H$ and
$H-K_S$ colors, and $K_S$ magnitudes, are derived from the measured
relative photometry by adopting the Two-Micron All
Sky Survey \citep[2MASS;][]{cutri_etal03} magnitudes for the primaries.  
The $K_S-L^\prime$ color of the companion to V522~Per was
calculated assuming $K_S-L^\prime=0.04$ for the F5~V primary 
\citep{bessell_brett88}.
A near-IR color-color diagram of the detected companions is
presented in Figure~\ref{fig_colors}.  Extinction corrections of
$A_J=0.087$, $A_H=0.055$, $A_K=0.033$, and $A_L=0.016$
have been applied to the colors of 
V522~PerB, based on $E(B-V)=0.10$ toward the
$\alpha$~Per cluster as measured by \citet{pinsonneault_etal98}.  We
have adopted $A_V=3.1 E(B-V)$, and the interstellar extinction law of
\citet*{cardelli_etal89}.  

Based on the colors and the $\Delta$$K_S$ magnitudes, we can infer
that HD~49197``C'' is a likely F--G background
star, whereas the remaining companions are consistent with
being late-type stars associated with their primaries.

\subsection{Astrometry}

Three of the four systems were observed at two different astrometric
epochs.  RX~J0329.1+0118 was observed only once, though prior-epoch astrometric 
data for it exist from \citet{sterzik_etal97}.  Table~\ref{tab_astrometry}
details for each binary the observed offset and position angle of the 
companion from the
primary during the first and second epochs of observation, as well as
the estimated offset and position
angle during the first epoch, had the system not been a common proper 
motion pair.  The
estimates are extrapolated backwards in time from the second-epoch 
astrometry, which for
all except HD~129333 was obtained with NIRC2 on Keck and was more accurate,
assuming proper motions from {\sl Hipparcos}
\citep{perryman_etal97} for HD~49197 and HD~129333, and from Tycho~2 
\citep{hog_etal00} for V522~Per and RX~J0329.1+0118.
Parallactic motions were also taken into account, as they are
significant for stars $\lesssim200$~pc at our astrometric
precision (several milli-arcsec).  The assumed proper motions and 
parallaxes for all stars are listed in Table~\ref{tab_properties}.
To ensure the correct propagation of
astrometric errors, the epoch transformations were performed
following the co-variant treatment of the problem, as detailed in 
\citet{lindegren97}.  

Proper motion diagrams for each object are presented
in Figures~\ref{fig_astrom_hd49197}--\ref{fig_astrom_rxj0329}. The
first- and second-epoch measurements are shown as solid points with
error bars, and the inferred first-epoch position (assuming
non-common proper motion) is shown only with errorbars.  The dashed line
reflects the combined proper and parallactic motion of the primary over the
period between the two epochs.
Below we discuss the evidence for common proper motion in each system.

\subsubsection{HD 49197}

\begin{figure}
\plotone{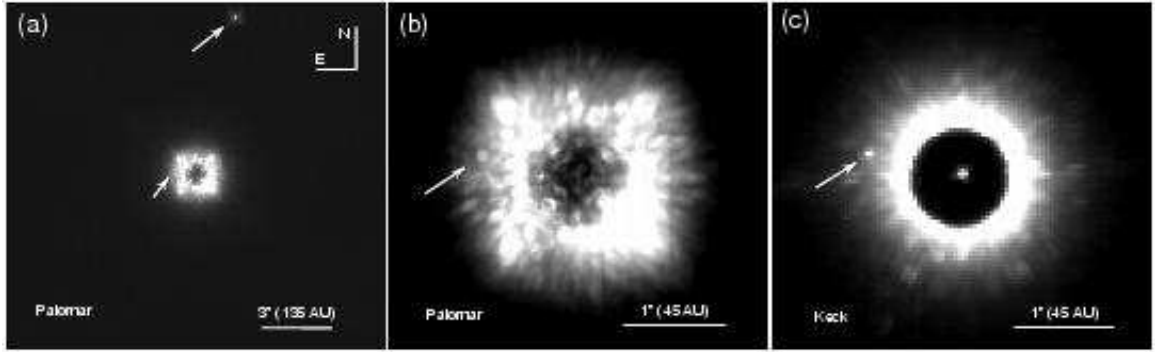}
\figcaption{$K_S$-band (2.15$\micron$) first (a, b) and 
second (c) epoch images of HD~49197.  Panel (a) shows both candidate
companions to HD~49197; panels (b) and (c) are zoomed in to
point out only the close-in bona-fide companion (HD~49197B).
The first-epoch image is the result of 24
median-combined 60~sec exposures with Palomar/PHARO, whereas the 
second-epoch image was
formed by median-combining six 60~sec exposures with Keck/NIRC2.  A
1.0$\arcsec$-diameter coronagraph occults the primary in both cases;
in the Keck image the coronagraph shows a residual 
$\approx$0.16\% transmission.  HD~49197B was initially unnoticed in
the first-epoch image, where its detection was hindered by the presence
of equally prominent AO speckles.
\label{fig_hd49197}}
\end{figure}

\begin{figure}
\plotone{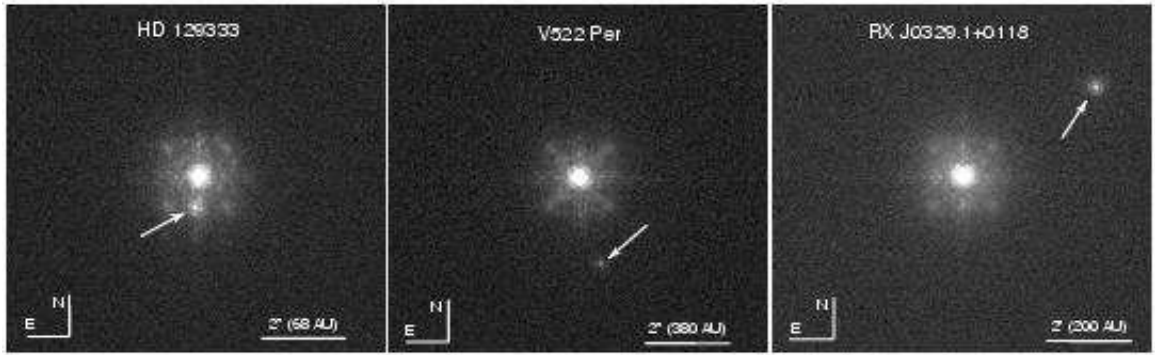}
\figcaption{Palomar images of the stellar companions.  The
HD~129333 image is taken through a narrow-band (1\%) Br$\gamma$ 
(2.166~$\micron$) filter, while the V522~Per and RX~J0329.1+0118 images are
taken at $K_S$.  Five dithered 1.4~sec exposures were aligned and 
median-combined to obtain each of the displayed images.
\label{fig_stellar}}
\end{figure}

\begin{figure}
\plotone{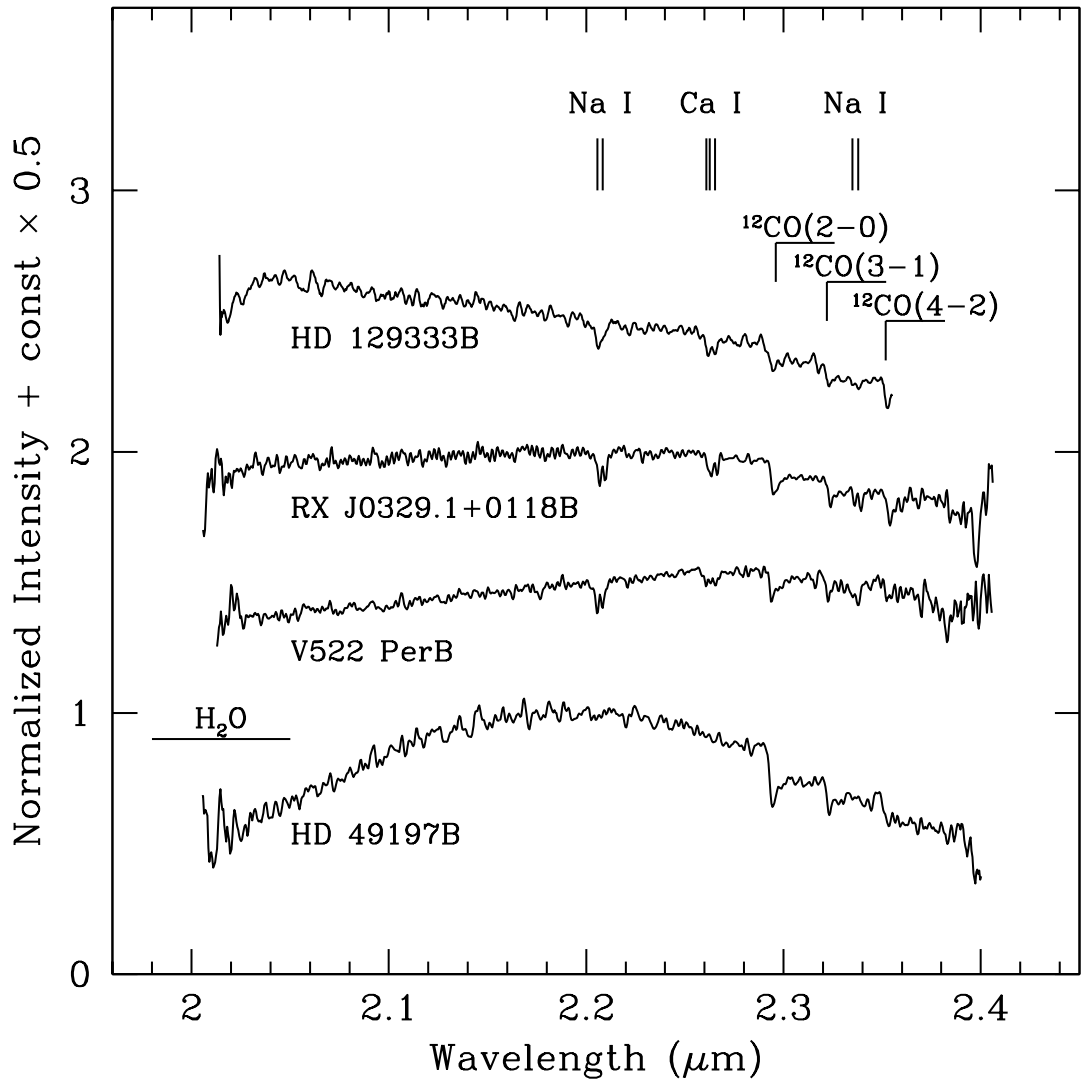}
\figcaption{$K$-band spectra of the low-mass companions 
from Palomar (HD~129333B; $R\approx1000$), and Keck (RX~J0329.1+0118B, 
V522~PerB, and HD~49197B; $R\approx2700$).  All spectra have been
normalized to unity at 2.20$\micron$ and offset by 0.5 in the vertical
axis.
\label{fig_Kspectra}}
\end{figure}

\begin{figure}
\plotone{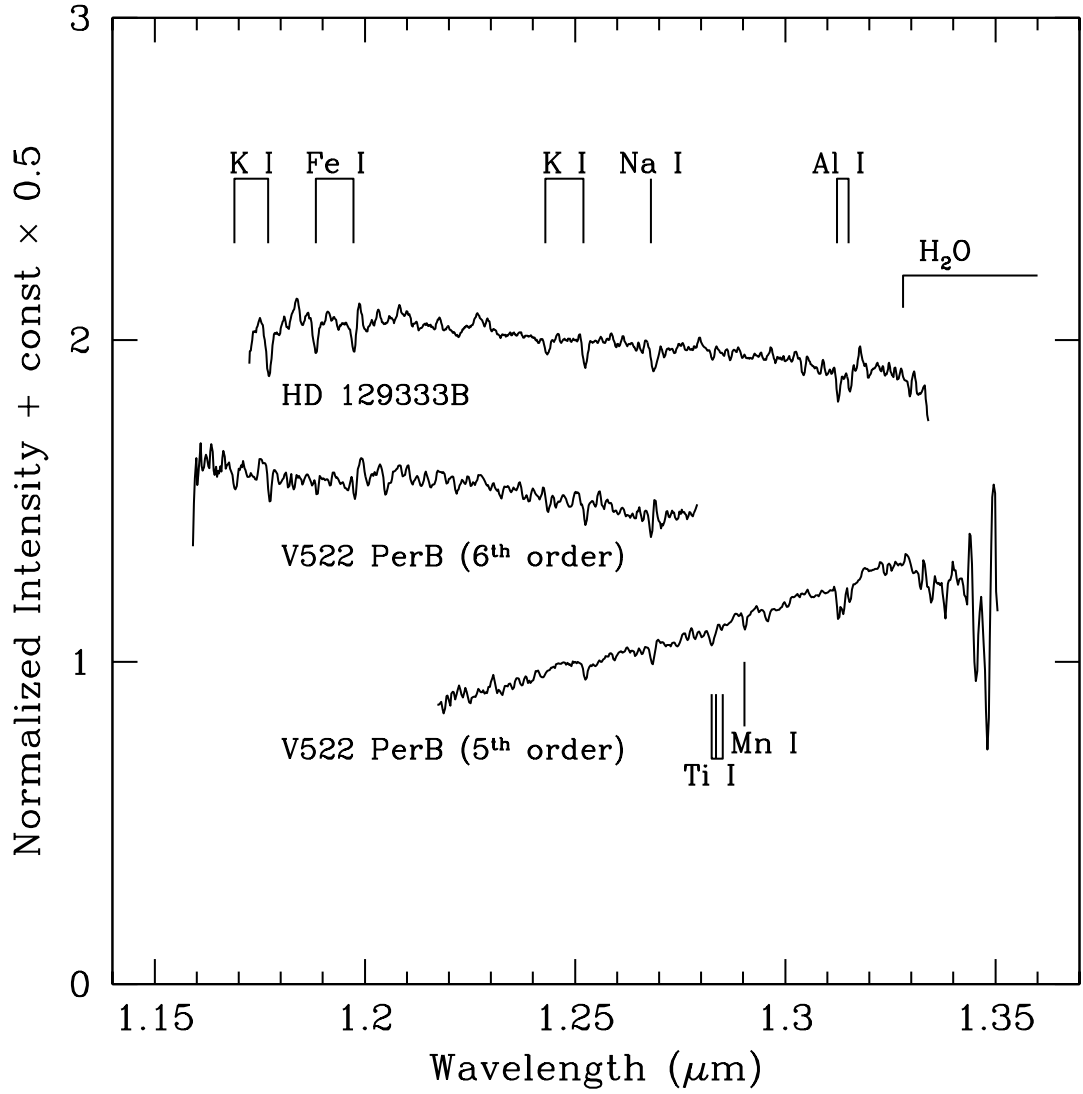}
\figcaption{$J$-band spectra of two of the low-mass companions from
Palomar (HD~129333B; $R\approx1200$), and Keck (V522~PerB;
$R\approx2400$ in the fifth order, and $R\approx2900$ in the sixth).
All spectra have been normalized to unity at 1.25$\micron$ and offset by
0.5 in the vertical axis.
\label{fig_Jspectra}}
\end{figure}

\begin{figure}
\plotone{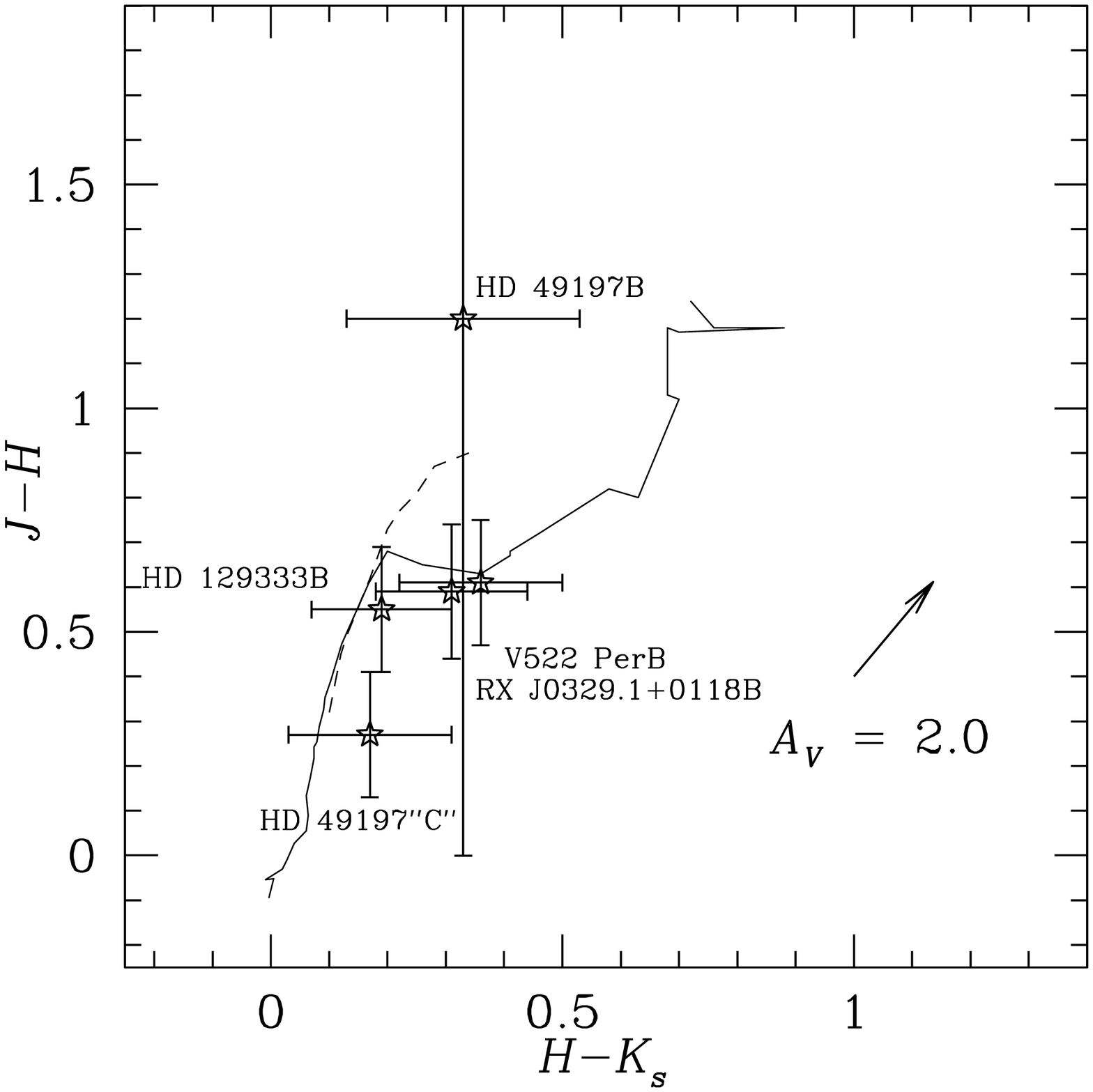}
\figcaption{Near-IR color-color diagram of the detected companions.  The
solid line represents the B2V--L8V main sequence, with
data compiled from \citet[][B2V--M6V]{cox00}
and \citet[][M8V--L8V]{kirkpatrick_etal00}.  The dashed line shows
the G0--M7 giant branch \citep{cox00}.  The \citeauthor{cox00} and the
\citeauthor{kirkpatrick_etal00} colors
are converted from the Johnson-Glass \citep{bessell_brett88} and 2MASS 
\citep{cutri_etal03} systems, respectively, to the CIT system using 
relations from \citet{carpenter01}.  HD~49197``C'' is too blue to be a
bona-fide low-mass companion to HD~49197.
The near IR colors of the remainder of the companions
agree well with their inferred spectral types (Table~\ref{tab_estimates}).
\label{fig_colors}}
\end{figure}

\begin{figure}
\plottwo{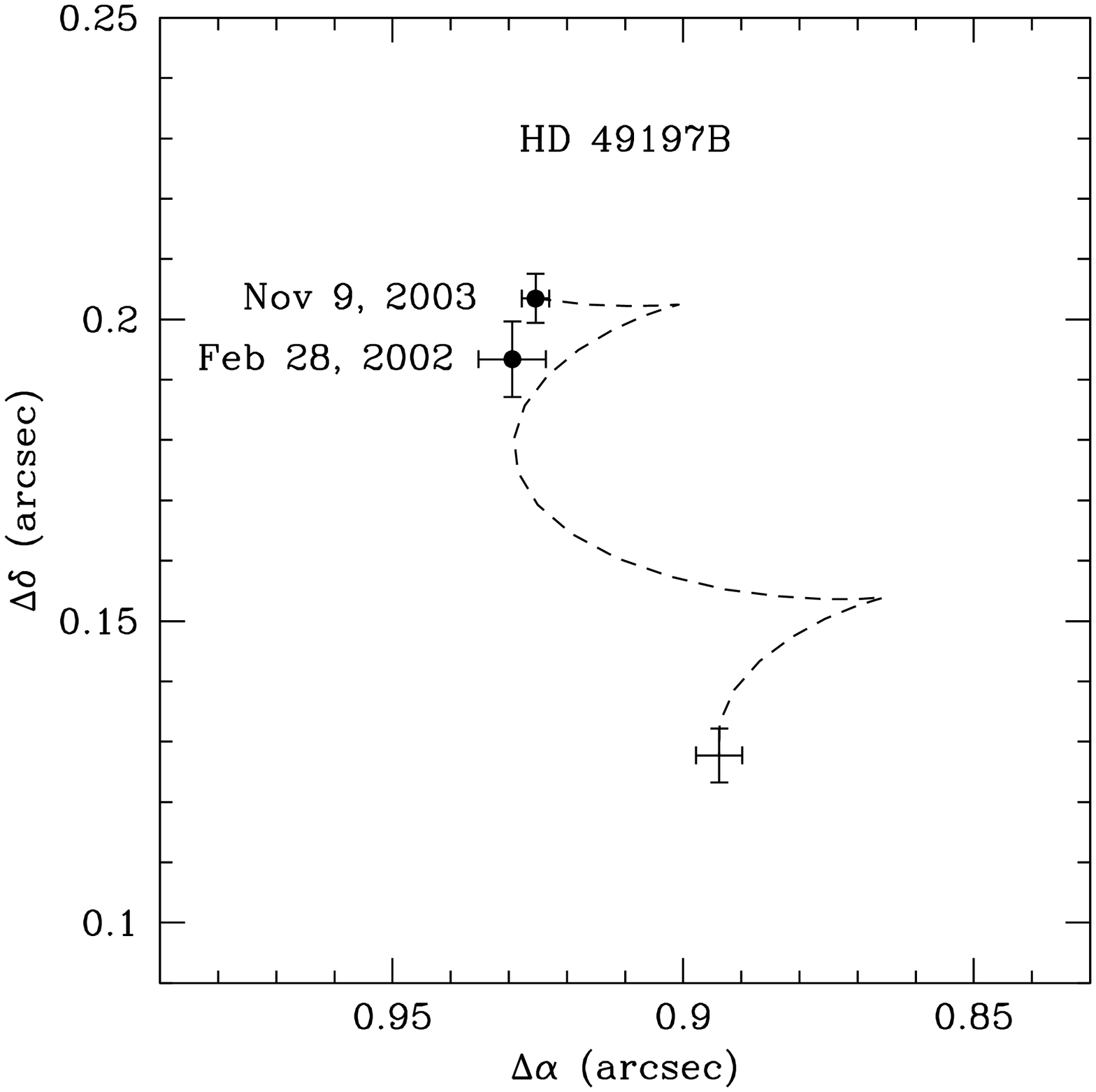}{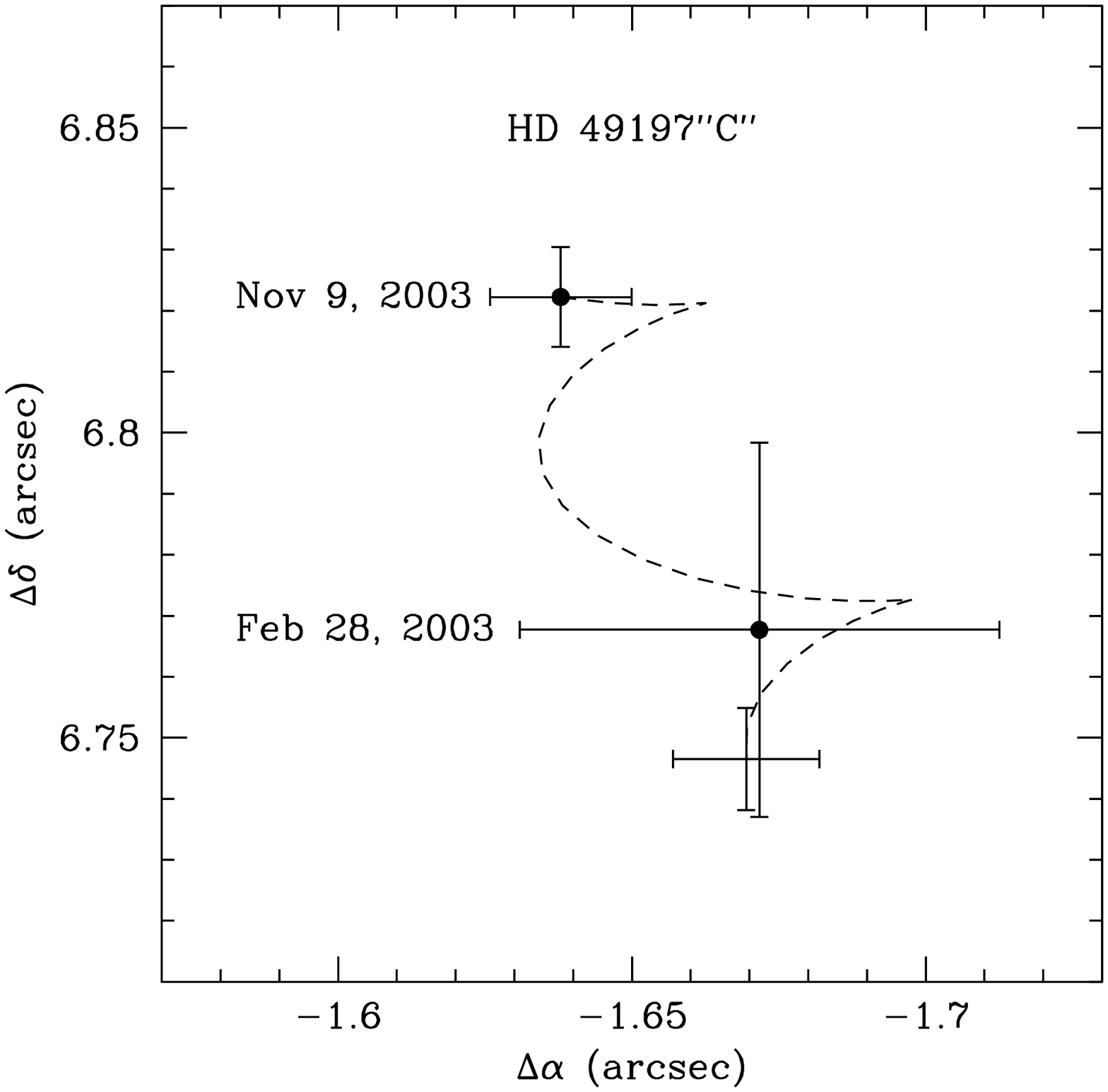}
\figcaption{Proper motion diagram for the two companions to
HD~49197.  The offsets from the primary at each observational
epoch are plotted as solid points with errorbars.  The inferred offsets
at the first epoch (assuming non-common proper motion) are shown just
with errorbars.  The dashed lines reflect the proper and parallactic
motions of the primary between the two epochs.  HD~49197B is a common
proper motion companion within the 1 sigma errors, while HD~49197``C'' is more
consistent with being a background object.
\label{fig_astrom_hd49197}}
\end{figure}

\begin{figure}
\plotone{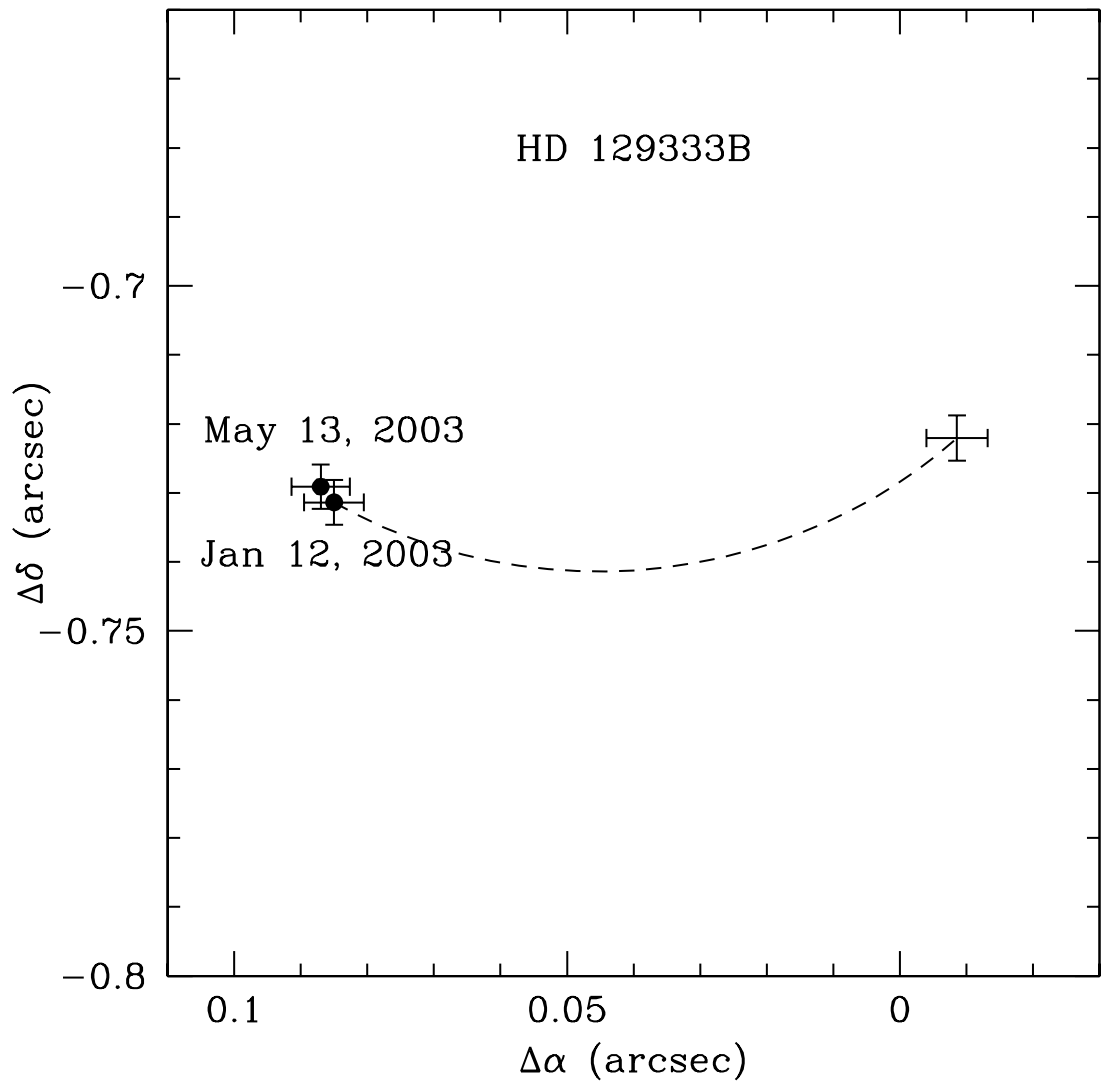}
\figcaption{Same as Figure~\ref{fig_astrom_hd49197} for the 
companion to HD~129333.  Within the 1 sigma errors, the companion shares the 
proper motion of the primary.
\label{fig_astrom_hd129333}}
\end{figure}

\begin{figure}
\plottwo{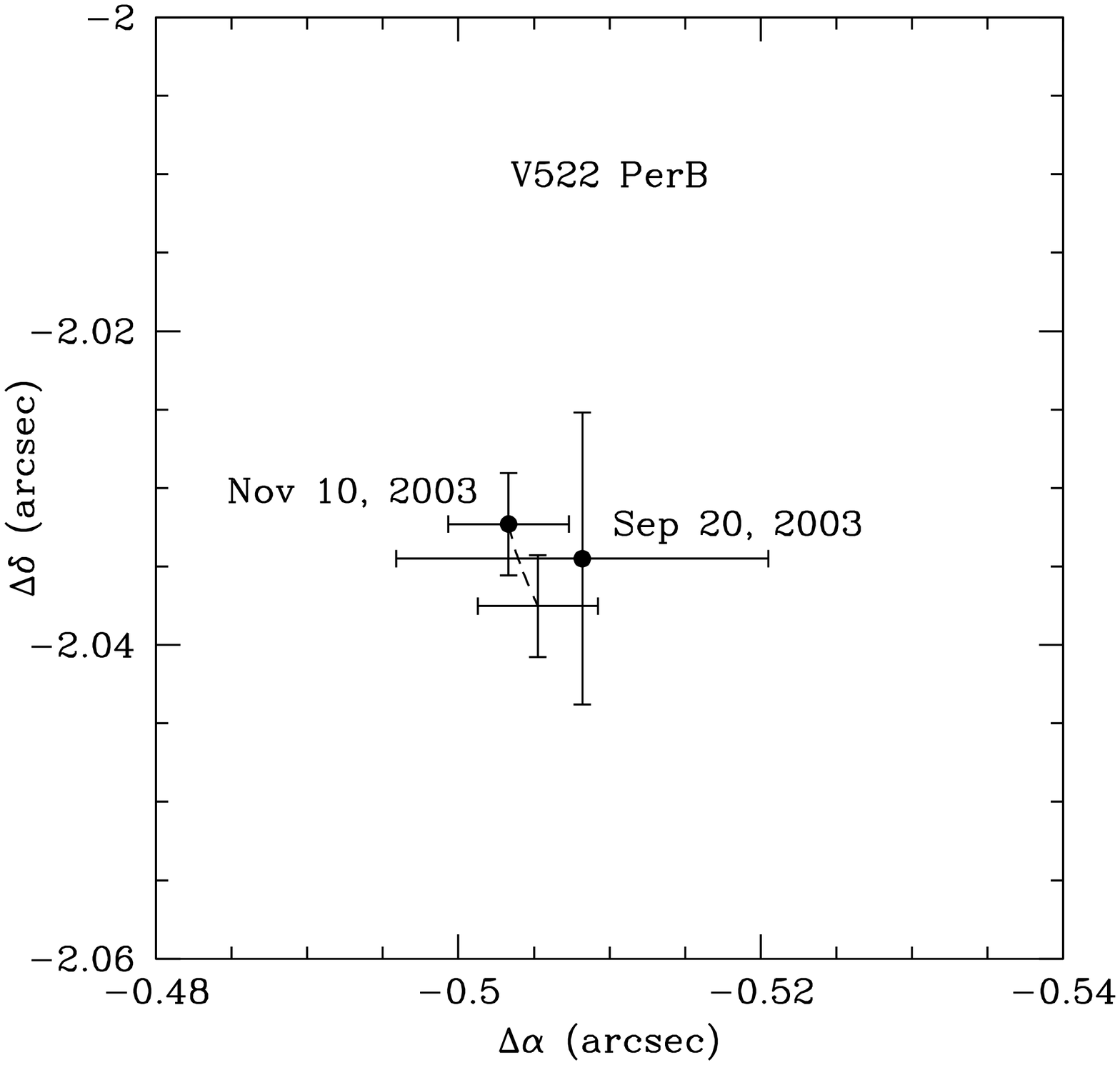}{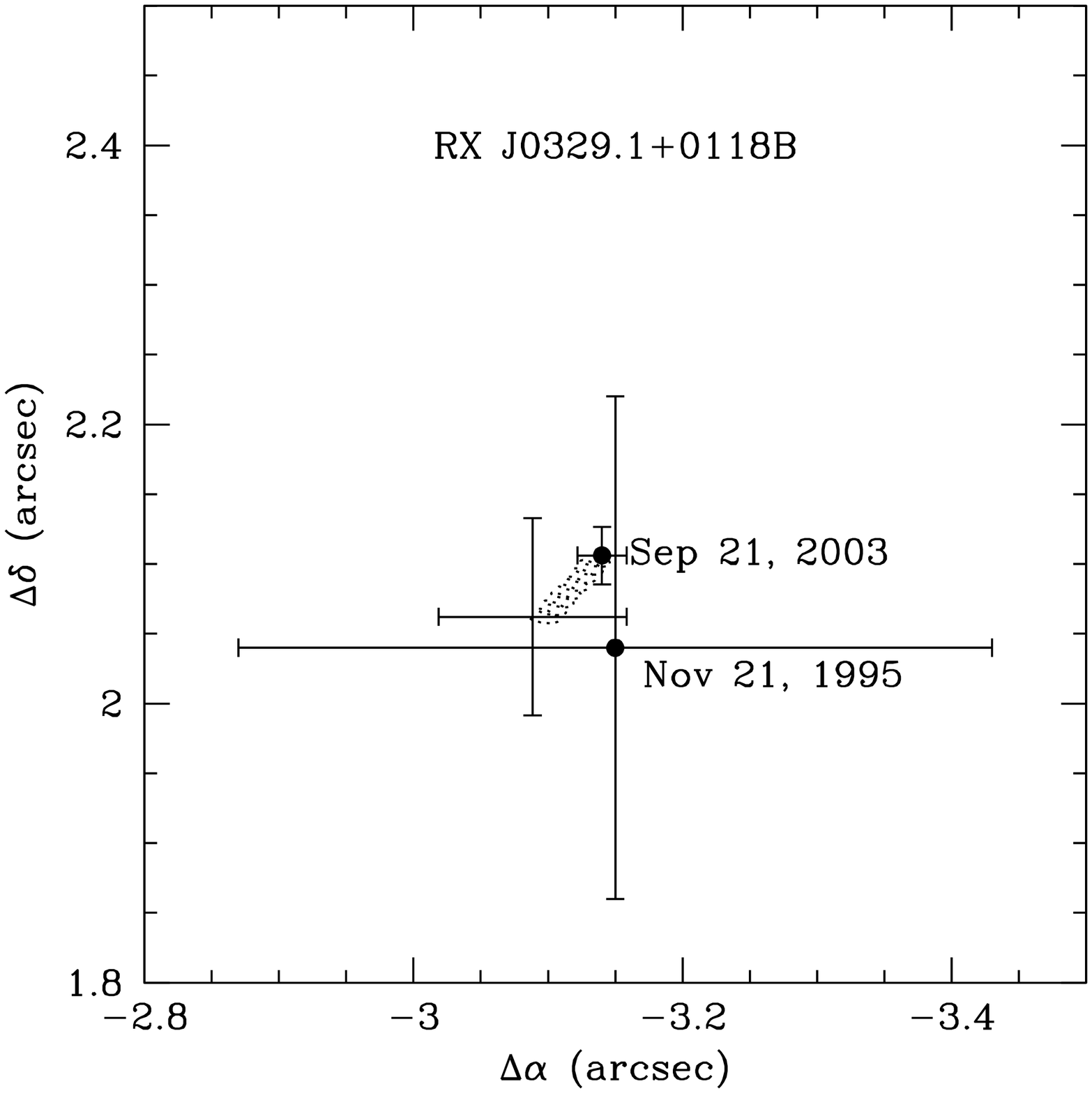}
\figcaption{Same as Figure~\ref{fig_astrom_hd49197} for the 
companions to V522~Per ({\it left}) and RX~J0329.1+0118 ({\it right}).  The Nov
21, 1995 data point for the companion to RX~J0329.1+0118 is from
\citet{sterzik_etal97} and is set to be at the mean epoch of their
observations (Nov 19--23, 1995).  The proper motion of the primaries 
between the observations epochs are too small to decide the physical
association of the companions within the astrometric errors.  The
probability of association in each system is estimated from near IR
spectroscopy (Section~\ref{sec_likelihood}).
\label{fig_astrom_rxj0329}}
\end{figure}

The existence of the
close (0.95$\arcsec$) companion to HD~49197 was unappreciated prior
to the second-epoch imaging: the star was followed up because of the more
distant (6.8$\arcsec$) candidate companion (HD~49197``C'').
Upon its discovery in the Keck image (Figure~\ref{fig_hd49197}c),
the close companion was recovered in the earlier Palomar image,
where a dark circular ring around the core at the radial distance 
of the first Airy null ($\approx$0.1$\arcsec$) distinguishes it from the 
telescopic speckles (Figure~\ref{fig_hd49197}b).

As is evident from Figure~\ref{fig_astrom_hd49197}, HD~49197B
(left panel) is much more
consistent with being a proper motion companion of the primary, 
than with being an unrelated background star projected along the same line of
sight, whereas the reverse holds for HD~49197``C'' (right panel).  
Therefore, we 
claim that the close-in HD~49197B is a bona-fide companion, whereas the
more widely-separated HD~49197``C'' is not. 

The astrometry for HD~49197B from the two observational epochs is not in 
perfect agreement, perhaps because of its orbital motion
around component A.  At a projected separation of 43~AU,
the orbital period will be $>$240~years \citep[assuming a circular face-on
orbit and a mass of 
1.16$M_\sun$ for the F5~V primary;][]{allendeprieto_lambert99}, 
resulting in a change in position angle of
$<$2.6$\degr$ between the two epochs.  This may explain the observed 
discrepancy of $0.64\degr\pm0.47\degr$ in position angle between the two 
observations, while the change in separation (2.4$\pm$5.8~mas) is
consistent with being zero.  
Future observations spanning a sufficiently long time-line
may be used to determine the dynamical mass of the HD~49197A/B
system.

\subsubsection{HD 129333}

The offset positions of the companion to HD~129333 in the two
observations taken 16 months apart are fully consistent with each other,
and inconsistent with the object being a background star
(Figure~\ref{fig_astrom_hd129333}).  HD~129333B
thus shares the proper motion of the primary, and is a bona-fide
companion.

\subsubsection{V522 Per and RX~J0329.1+0118}

The astrometry for these two systems (Figure~\ref{fig_astrom_rxj0329})
is inconclusive because of their smaller proper motions, insufficient
time-span between the observations, and/or less accurate astrometry.  
The likelihood of physical
association of the companions with the primaries is 
investigated from low-resolution spectroscopy in
Section~\ref{sec_likelihood} below.

\subsection{Spectroscopy \label{sec_spectroscopy}}

Infrared spectral classification of M--L dwarfs is done most
successfully in the $J$ and $H$ bands, where a suite of indices based on
the relative strengths of H$_2$O, FeH, \ion{K}{1}, and \ion{Na}{1} 
absorption have been developed to characterize their effective temperatures
\citep[\citealt*{slesnick_etal04};][]{geballe_etal02, gorlova_etal03, 
mclean_etal03} and surface gravities 
\citep{gorlova_etal03}.  However, 
spectroscopy of cool companions in the bright halos of their primaries
is often more difficult at $J$ and $H$ than at $K$, because of larger flux
contrast and poorer quality of the AO correction at shorter
wavelengths.  Here we present $J$-band spectra of the two brighter
companions: HD~129333B, and V522~PerB.  Cool dwarfs can be
classified at $K$ band based on the strength of their H$_2$O absorption
shortward of 2.05$\micron$, the depth of the CO 2.29$\micron$
bandhead, and the equivalent width (EW) of the \ion{Na}{1} 2.21$\micron$
doublet \citep{kleinmann_hall86, ali_etal95, mclean_etal03,
slesnick_etal04}.  Even cooler, late-L and T dwarfs, are
best characterized by the strength of their CH$_4$ absorption at 
$>$2.20$\micron$ \citep{burgasser_etal02, mclean_etal03}.  We present
$K$-band spectra of all four companions discussed in this
paper.

Spectral types of the detected companions have been determined following
infrared absorption line classification systems at $K$-band 
(for \ion{Na}{1}, \ion{Ca}{1}, CO) from
\citet[][spectral types M6V--T8]{mclean_etal03} and
\citet[][F3V--M6V]{ali_etal95}, and at $J$-band (for \ion{K}{1} and
\ion{Ti}{1}) from
\citet[][M4V--L8]{gorlova_etal03} and \citet[O7--M6]{wallace_etal00}.  
Luminosity classes are based on the relative strengths of \ion{Na}{1}, 
\ion{Ca}{1}, and CO
absorption at $K$ band \citep{kleinmann_hall86}, and on the strengths of
\ion{Mn}{1}, and \ion{K}{1} absorption at $J$
\citep{wallace_etal00, gorlova_etal03}.

We have avoided the use of temperature-sensitive H$_2$O
indices that span large fractions of the spectrum ($\gtrsim$4\% total band
width, e.g., the H$_2$OD index of \citet{mclean_etal03}, or the H$_2$O-2
index of \citet{slesnick_etal04}) because of their strong dependence on 
the overall continuum shape of the 
spectrum, and because of the uncertainties in the spectral shapes of faint
companions extracted from the halos of bright objects
(Section~\ref{sec_obs_spec}).  We have used the $J$-band 1.34~$\micron$ 
H$_2$O index of \citet{gorlova_etal03} however, which measures only the
onset of H$_2$O absorption at that wavelength, and is narrow (1.4\% band width).

Below we analyze the $K$- and $J$-band spectra of the
companions to HD~49197, HD~129333, V522~Per, and RX~J0329.1+0118.
The inferred effective temperatures
and spectral types are provided in Tables~\ref{tab_eqw} and
\ref{tab_estimates}.

\subsubsection{$K$-band Spectroscopy of HD~49197B
\label{sec_spec_hd49197b}}

Our spectrum of HD~49197B (Figure~\ref{fig_Kspectra}) shows strong CO 
and H$_2$O bands characteristic of ultra-cool dwarfs, but lacks a CH$_4$
absorption feature, indicating that it is earlier than L8
\citep{geballe_etal02}.  On the other hand, \ion{Na}{1} absorption
is also absent from the spectrum, pointing to a spectral
type of L3 or later \citep{mclean_etal03}, independent
of gravity \citep[cf.\ Figure~8 in][]{gorlova_etal03}.  
Following the analysis of
$R$$\sim$2000 $K$-band spectra of M6--T8 dwarfs in \citet{mclean_etal03}, 
we form a CO absorption index from
the ratio of the median flux in an absorbed (2.298--2.302~$\micron$)
vs.\ an unabsorbed (2.283--2.287~$\micron$) region of the 
spectrum.  We find CO=0.80$\pm$0.03
(where the error has been estimated as the quadrature sum of the
relative uncertainties of the median in the two spectral regions), which 
indicates
a spectral type of L7 or later.  However, the CO index is not very
sensitive to L-dwarf temperatures, and varies by up to 0.15
(60\% of its total range of variation between spectral types M5 and T2)
within the same spectral type \citep[Figure~13 in][]{mclean_etal03}.

Alternatively, we can also use the absolute $K_S$-band magnitude of 
HD~49197B to
estimate its spectral type, following the empirical relation of
\citet{kirkpatrick_etal00}, based on a sample of 24 M and L dwarfs with
measured parallaxes:
\begin{equation}
M_{K_S} = 10.450 + 0.127({\rm subclass}) + 0.023({\rm subclass})^2,
\end{equation}
where subclass~= $-1$ for M9~V, 0 for L0~V, 1 for L1~V, etc.  The
scatter about the fit is approximately 1 subclass.  From the inferred
apparent $K_S$ magnitude of HD~49197B, and from the parallax of the
primary, we obtain $M_{K_S}=11.04\pm0.24$ for the secondary, which
corresponds to a spectral type of L3$\pm$1.5.

We assign a final spectral type of L4 with an uncertainty of 1 subclass.
This is based on the intersection of the results from our spectroscopic
analysis, suggesting L3--L7, and from the $K_S$-band absolute magnitude, 
pointing to L1.5--L4.5.  A spectral type of L4$\pm$1 for HD~49197B is also 
consistent with a by-eye comparison of its $K$-band spectrum with the 
grid of L-dwarf standards from \citet{leggett_etal01}.

\subsubsection{$K$-band Spectroscopy of HD 129333B, RX~J0329.1+0118B, 
and V522~PerB \label{sec_kband_spec2}}

We classify the $K$-band spectra of these companions following 
the analyses of \citet{ali_etal95} and \citet{kleinmann_hall86}, whose
data are taken at similar resolutions to ours ($R$=1380--3900 and 2500--3100,
respectively), and span the F8--M7 spectral type range.  We employ the spectral 
classification sequence of \citeauthor{ali_etal95}, whose empirically-derived
indices are based on a larger sample of dwarf stars than in
\citeauthor{kleinmann_hall86}.

Our reduced spectra were first shifted to 0~km~s$^{-1}$ heliocentric velocity, 
where the shift was determined by fitting Gaussian profiles 
to the \ion{Na}{1} doublet, and comparing the fitted line centers to
their values in the solar spectrum \citep{mohler55}.  For consistency
with the \citet{ali_etal95} spectral classification, we 
have chosen the same bands for integrating the \ion{Na}{1}
(2.21$\micron$) \ion{Ca}{1} (2.26$\micron$), and $^{12}$CO(2--0) 
(2.29$\micron$) absorption.  The continuum in the target spectra was
fit to three regions devoid of absorption lines:
2.0907--2.0951$\micron$, 2.2140--2.2200$\micron$, and 2.2873--2.2925$\micron$.
These have been selected as a
combination of the continuum regions used by \citet{ali_etal95} and
\citet{kleinmann_hall86}, so as to constrain the fit on both sides
of the \ion{Na}{1} doublet \citep[as in][where the continuum-fitting 
regions are widely-separated from the \ion{Na}{1} 
lines]{kleinmann_hall86}, as well as near it \citep[as in][where the
continuum is constrained only on the long-wavelength side of the
\ion{Na}{1} lines]{ali_etal95}.  

The absorption strength in each band was obtained as an equivalent width, 
by integrating the profile of the spectrum in the band
with respect to a global continuum level defined by the
three continuum bands.  The only exception is
the EW of the $^{12}$CO(2--0) first overtone bandhead, for which we
have adopted the mean continuum level of the third continuum band
(as in \citealt{kleinmann_hall86} and \citealt{ali_etal95}).
The one-sigma errors in the EWs were calculated by
propagating the r.m.s.\ noise of the spectrum in the nearest continuum
band, assuming independent pixel variances.  
The EWs of \ion{Na}{1} and \ion{Ca}{1} were
corrected for corresponding absorption in the spectra of the telluric
standards, the EW of which \citep[0.84$\pm$0.57~\AA\ for \ion{Na}{1} and
1.14$\pm$0.36~\AA\ for \ion{Ca}{1} in G 
stars;][]{ali_etal95} was added to that measured for the 
companions.  The final EWs are listed in Table~\ref{tab_eqw}.

To infer effective temperatures for the companions to 
HD~129333, RX~J0329.1+0118, and V522~Per, we employ empirical relations 
between the strength of $K$-band
\ion{Na}{1}, \ion{Ca}{1}, and CO absorption and effective temperature 
($T_{\rm eff}$), as determined by \citet[][their Table~4]{ali_etal95}. 
Na and CO produce the most characteristic $K$-band
features of cool stars.  However, their absorption strengths are inaccurate
tracers of temperature for stars cooler than 4000~K, and are, in
addition, gravity-sensitive \citep{kleinmann_hall86,gorlova_etal03}.
On the other hand,
\ion{Ca}{1} transitions in the $K$-band require the population of higher
energy states, and hence are more temperature-sensitive. However, their
absorption strength is degenerate with the stellar effective temperature:
\ion{Ca}{1} absorption at 2.26$\micron$ peaks at $\sim$3500~K, and 
decreases for higher, and lower effective temperatures 
\citep{kleinmann_hall86, ali_etal95}. This behavior is fit via two separate 
$T_{\rm eff}$ vs.\ EW(\ion{Ca}{1}) relations in \citet{ali_etal95}.
By combining the information from \ion{Ca}{1} with that
from \ion{Na}{1} and $^{12}$CO(2--0), we can break this degeneracy, and
use the more temperature-sensitive \ion{Ca}{1} index to
constrain the effective temperature for each star to 
within $\sim$300~K \citep[the quoted uncertainty of the \ion{Ca}{1}
index in][]{ali_etal95}.  For HD~129333B, with a \ion{Ca}{1} absorption
strength near the
breaking point between the ``hot'' and ``cool'' relations, we take the
average of the two estimates.  For V522~PerB and RX~J0329.1+0118B,
whose \ion{Na}{1} and CO absorption is indicative of temperatures
$T_{\rm eff}<3400$~K, we use only the ``cool'' relation.  The effective
temperatures inferred from \ion{Ca}{1} absorption are listed in
Table~\ref{tab_eqw}.  As in \citet{ali_etal95}, we adopt a spectral 
type vs.\ effective
temperature classification for M dwarfs from \citet{bessell91}, and obtain
spectral types of M1, M3, and M4 for HD~129333B, RX~J0329.1+0118B and 
V522~PerB, respectively, with an uncertainty of 2 spectral subtypes.
A visual inspection and comparison of the strengths of the various 
absorption features with $K$-band spectral sequences
from \citet{leggett_etal01} and \citet{wallace_hinkle97} confirms these
results.  Given the comparable values of their \ion{Na}{1} and
$^{12}$CO(2--0) EWs, all stars have likely dwarf 
gravities \citep[cf.\ Figure~7 in][]{kleinmann_hall86}.

\subsubsection{$J$-band Spectroscopy of HD 129333B and V522~PerB
\label{sec_jband_spec}}

We further constrain the spectral types of HD~129333B and of V522~PerB
from their $J$-band spectra.  These 
show the characteristic absorption features of M stars: \ion{K}{1} lines
at 1.169$\micron$, 1.177$\micron$, 1.243$\micron$, and 1.252$\micron$,
\ion{Fe}{1} at 1.188$\micron$ and 1.197$\micron$, \ion{Na}{1} at
1.268$\micron$, \ion{Al}{1} at 1.312$\micron$ and 1.315$\micron$, and
H$_2$O absorption at $\lambda>1.34\micron$ (Figure~\ref{fig_Jspectra}).  
The spectrum of V522~PerB exhibits also
\ion{Ti}{1} and \ion{Mn}{1} absorption over 1.282--1.298~$\micron$.
Spectral classification at $J$ band was done based on the depth of the 
H$_2$O and \ion{K}{1} absorption, following the index definitions of 
\citet{gorlova_etal03}.  
After re-sampling our data to match the $R$$\approx$350 spectral
resolution of \citeauthor{gorlova_etal03},
we form the 1.34$\micron$ water index as the ratio of the mean
fluxes (in a 0.004~$\micron$ wide region) at 1.336$\micron$ and 
at 1.322$\micron$, and we
measure the \ion{K}{1} EW by integrating the absorption over the region
1.2375--1.2575$\micron$.  Because \ion{K}{1} absorption in the solar-like
photospheres of the telluric standards is small
\citep[$EW$(\ion{K}{1}~$\lambda1.14\micron) \approx 
0.1 \times EW$(\ion{Na}{1}~$\lambda2.21\micron) = 0.08$~\AA\ for the 
Sun;][]{mohler55,ali_etal95}, the \ion{K}{1} EW measurements of the
companions were not corrected for it.

For HD~129333B, we cannot estimate the strength 
of the 1.34$\micron$ water absorption because of insufficient spectral
coverage.  The EW of \ion{K}{1} indicates a spectral type of 
M2--M4.  Averaging this with our $K$-band estimate of M1$\pm$2, we
assign a spectral type of M2$\pm$1 for HD~129333B.
Given the youth of HD~129333 ($<$120~Myr; see
Section~\ref{sec_masses}), the companion may
have lower-than-dwarf gravity.  The effect of this on alkali
absorption lines in the near IR is degenerate with temperature
\citep{gorlova_etal03, mcgovern_etal04}, and could be compensated by a
later spectral type.  However,
spectral types later than M3 are inconsistent with the
depth of the strongly temperature-sensitive \ion{Ca}{1} absorption
in this star (Section~\ref{sec_kband_spec2}).  In addition, HD~129333B 
lacks noticeable \ion{Mn}{1} absorption, which is weak in M dwarfs but
grows deeper with decreasing surface gravity in M stars
\citep[Fig.~9 in][]{wallace_etal00}.  Hence, we conclude that 
HD~129333B is a M2$\pm$1 dwarf.

For V522~PerB, both
the \ion{K}{1} EW and the water absorption index point to a spectral
type of M3--M5, in agreement with our $K$-band estimate (M4$\pm$2).  The
stronger \ion{Ti}{1} absorption than in HD~129333B is also consistent 
with a cooler photosphere.  We
thus assign a spectral type of M4$\pm$1 to V522~PerB.  The strength of
the \ion{Mn}{1} transition indicates a potentially
subgiant surface gravity, though as noted in
Section~\ref{sec_kband_spec2}, the effect is not seen at $K$ band.

\section{DISCUSSION}

\subsection{Likelihood of Physical Association \label{sec_likelihood}}

Our astrometric follow-up of HD~49197B and HD~129333B confirmed common
proper motion between these two companions and their respective
primaries.  However, the smaller proper motions of V522~Per and RX~J0329.1+0118
prevented us from concluding the same for their respective 
companions, given the time-span of our observations.  The
probability of physical association in these systems can be inferred
from the spectroscopically determined spectral types and absolute
magnitudes of the companions.  If the absolute magnitude inferred
from the spectral type of a 
companion agrees with its measured apparent magnitude at the heliocentric 
distance of the respective primary, then the companion is likely to be a
bona-fide one (modulo the space density of stars of similar spectral
type as the companion).  

\begin{figure}
\plotone{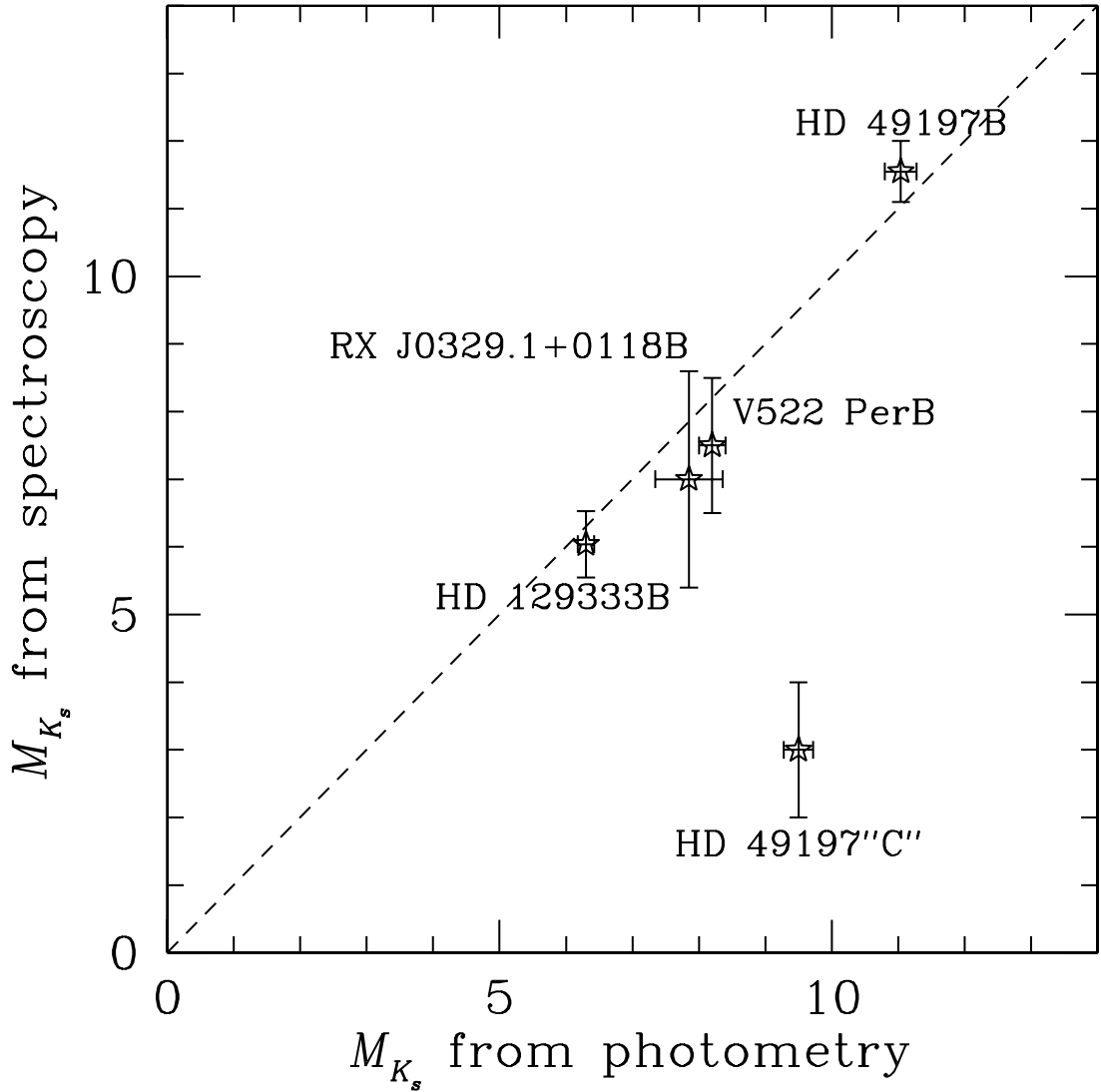}
\figcaption{Comparison of the photometrically derived absolute
$K_S$-band magnitudes of the companions (assuming the 
heliocentric distances of the corresponding
primaries) to the spectroscopically inferred ones.  The dashed line has a
slope of unity.  The location of HD~49197``C'' (of which no spectra were
taken) along the vertical axis is based on a spectral type (F--G)
inferred from its near IR colors (Section~\ref{sec_anal_photometry}).  
HD~49197``C'' is intrinsically too bright to be at the same heliocentric
distance as HD~49197, whereas the remainder of the companions are
consistent with being at the distances of their respective primaries.
\label{fig_absolute_mags}}
\end{figure}

Figure~\ref{fig_absolute_mags} presents a 
comparison of the spectroscopic vs.\ photometric absolute magnitudes.
The correspondence is good for the four companions followed up
via near IR spectroscopy, indicating that 
they are at similar heliocentric distances as their primaries, and are
thus likely to be physically bound to them.  The location of the
remaining candidate companion (HD~49197``C'')
along the ordinate is inferred from its near-IR colors.  
As mentioned in Section~\ref{sec_anal_photometry}, HD~49197``C'' is a 
likely F--G star \citep[$2.0 \leq M_K \leq 4.0$;][]{cox00}, i.e., it is
intrinsically too bright to be associated with HD~49197 (F5~V) given its 
faint apparent magnitude.

A robust statistical analysis of the likelihood of chance alignment in
the four systems discussed here is not yet possible at this stage of the
survey.  They are only a fraction of the ones discovered to have
candidate companions.  With follow-up observations still
in progress, the exact number of bound systems is unknown.  
We defer a discussion of the
companion chance alignment probability based on the full
ensemble statistics until a later paper.  Here we consider these
probabilities only on a per system basis.  To give an approximate idea
of the limited statistics from which these preliminary results are
extracted, we point out that to date we have analyzed multi-epoch
astrometric data for approximately 15 stars (mostly from the deep
survey) with faint ($\Delta K_S>3$, i.e., expected to be of
spectral type M or later) companions within 4$\arcsec$.  

We base our calculation of the probability of false association in each 
system on the empirically determined spatial density
of cool objects (M--T spectral types) in the solar neighborhood.  
There are 112 such known objects in the northern ($\delta>-20\degr$) 8~pc 
sample \citep{reid_etal03}.  The northern 
8~pc sample covers 65\% of the sky, and is estimated to be $\sim$15\%
incomplete.  The total number of cool
objects and white dwarfs within 8~pc of the Sun is therefore expected to
be 198, with a volume density of 0.10~pc$^{-3}$.  This estimate is based 
on a small fraction of the 
thin disk population \citep[scale height 325~pc;][]{bahcall_soneira80}
of the Galaxy and hence should not vary substantially as a
function of galactic latitude.  

We then calculate the number of cool dwarfs expected to be
seen in projection toward each system within a conical volume of radius
4$\arcsec$ centered on the star, with the observer at the apex of the
cone.  We truncate the radial extent of the conical volume by requiring 
that the apparent
$K$ magnitude of a projected companion falls within the limits allowed
by the spectral type (and hence, absolute magnitude) of the detected
one.  Absolute $K$ magnitudes for the M2--4 dwarfs discussed here have 
been adopted from \citet{bessell91}.  Although
\citeauthor{bessell91}'s M dwarf classification system pre-dates the
discovery of ultra-cool dwarfs (later than M5), it remains 
valid for early M dwarfs.  For L4$\pm$1 spectral types we adopt absolute
$K$ magnitudes from \citet{dahn_etal02}.

The expected number $\mu$ of unrelated cool dwarfs 
within the relevant volume around each star is listed in the last
column of Table~\ref{tab_estimates}.  Given
that for all stars $\mu\ll 1$, we can assume that the event of seeing an
unrelated field object in the vicinity of one of our program stars
is governed by Poisson statistics.  Hence, the
probability of finding one or more such dwarfs near any given star (i.e.,
the probability of chance alignment) is $1-e^\mu \approx
\mu$.  As seen from Table~\ref{tab_estimates}, after having followed
up the companions spectroscopically, we can claim with $\geq$99.99\%
certainty in each case that the companion is physically associated with its 
respective primary.  As discussed above, such probabilities need
to be regarded in the context of the ensemble statistics.  Within our
sample of 4 spectroscopically confirmed companions, the probability that
at least one is a false positive is $3\times10^{-3}$.
This exemplifies the power of spectroscopic
follow-up as an alternative to multi-epoch astrometry in constraining the 
likelihood of physical association in a system.

\subsection{Stellar Ages and Companion Masses \label{sec_masses}}

We estimate the ages of the primaries (Table~\ref{tab_properties}) from
published data on their chromospheric activity, \ion{Li}{1} equivalent
width, and kinematic association with young moving groups.
Masses for each of the companions (Table~\ref{tab_estimates}) were 
determined either from the low-mass pre-main 
sequence evolutionary models of \citet{baraffe_etal98}, or from the
brown-dwarf cooling models of \citet[][``DUSTY'']{chabrier_etal00} and 
\citet{burrows_etal97}.  We have not used 
the dust-free ``COND'' models of \citet{baraffe_etal03}, since they are more
appropriate for temperatures $\lesssim$1300~K (i.e., for T dwarfs) when 
all grains are expected to have gravitationally settled below the photosphere.

\paragraph{HD 49197B.} 
From the strength of Ca~H \& K core emission in
Keck/HIRES spectra of the primary, \citet{wright_etal04} determine an
age of 525~Myr for HD~49197, which we assume accurate to within 
$\approx$50\%, 
given the variation in chromospheric activity of solar-type stars
\citep{henry_etal96}.  No other age-related indicators
exist in the literature for this F5 star.  From our own high-resolution
optical spectra, we measure EW(Li~$\lambda6707.8$)~=
80~m\AA\ (Hillenbrand et al., in prep.), consistent with a
Pleiades-like \citep*[120~Myr;][]{stauffer_etal98} or older age.
Hence, we adopt an age of
260--790~Myr for the primary.  Assuming co-evality, the mass of the
secondary is $0.060_{-0.025}^{+0.012} M_\sun$ \citep{burrows_etal97,
chabrier_etal00}, where the range of masses accomodates the one
sigma error in the inferred absolute magnitude of the secondary, and 
the allowed age range for the primary.  HD~49197B is thus a brown
dwarf.

\paragraph{HD 129333B.}
The primary is a well-known young star, kinematically belonging to
the Local Association \citep[Pleiades moving group, 
20--150~Myr;][]{soderblom_clements87,montes_etal01a}.  Results from the Mount
Wilson spectroscopic survey \citep{soderblom85} and from the Keck/Lick 
r.v.\ program \citep{wright_etal04} show strong \ion{Ca}{2} H \& K emission,
indicating high levels of chromospheric activity and youth.
\citeauthor{wright_etal04} list an age of $<$10~Myr for this star,
though the chromospheric activity-age relation is not reliable for
stars that young, in part because of the large variance in rotation rates of
stars younger than 50--80~Myr \citep[e.g.,][]{soderblom_etal93}, and 
because the relation is not calibrated at such young ages.
\citet{montes_etal01b} report strong \ion{Li}{1} absorption 
($EW$(\ion{Li}{1})~= 198~m\AA), and conclude that the star is
``significantly younger'' than the Pleiades
\citep[120~Myr;][]{stauffer_etal98}.  Assuming an age of 10--100~Myr for
the system, we estimate the mass of the secondary at $0.20_{-0.08}^{+0.30}
M_\sun$ \citep[from models of][]{baraffe_etal98}.

\paragraph{V522 PerB.}
The primary is a member of the $\alpha$~Per open cluster, confirmed by
photometry, kinematics, and spectroscopy \citep{prosser92}.  From
high-resolution spectroscopy and determination of the lithium depletion
boundary in the cluster, \citet{stauffer_etal99} determine an age of
90$\pm$10~Myr, consistent with a recent age estimate from upper
main-sequence turn-off fitting \citep{ventura_etal98}.  Using the Lyon 
group stellar evolution models \citep{baraffe_etal98}, we determine a mass of
0.085--0.15~$M_\sun$ for the secondary.  However, from their sub-stellar 
``DUSTY'' code \citep{chabrier_etal00}, the treatment of dust opacity in
which may be more appropriate for this cool ($\sim$3200~K) star, we 
find that its mass is
$\geq$0.10~$M_\sun$.  We thus estimate 0.10--0.15~$M_\sun$ for the mass
of V522~PerB.

\paragraph{RX~J0329.1+0118B.}
\citet{neuhauser_etal95} list RX~J0329.1+0118 (G0~IV) as a 
fast-rotating ($v \sin
i=70$~km/s) X-ray source south of Taurus, with a \ion{Li}{1} equivalent width
of 110~m\AA: all indicators of relative youth.  Assuming a common origin
and distance with the stars in the Taurus molecular cloud 
\citep*[140~pc;][]{kenyon_etal94}, the authors claim that its
bolometric luminosity is higher than that of a zero-age main-sequence
star, and the star is therefore likely in the pre-main sequence (PMS)
phase.  From a proper-motion survey of PMS stars in Taurus-Auriga however,
\citet{frink_etal97} find that the young stars south of Taurus discussed
in \citet{neuhauser_etal95} are kinematically unrelated to those in Taurus,
and that star formation in the two complexes must have been triggered by
different events.  From the Pleiades-like \ion{Li}{1} equivalent width
of RX~J0329.1+0118, we assign an age of $\approx$120~Myr for
this star.  Given the spectral type of the secondary, its
mass is $0.20_{-0.10}^{+0.30} M_\sun$ \citep{baraffe_etal98}.

\subsection{HD 129333: Binary or Triple?}

The existence of a stellar companion to HD~129333 has already been
inferred in the r.v.\  work of \citet[][DM91]{duquennoy_mayor91}, who
find that the star is a long-period single-lined spectroscopic binary
(SB1).  
From their derived orbital parameters, the authors determine a secondary
mass $M_2\geq0.37M_\sun$, and suggest that the star
be targeted with speckle interferometry to attempt to resolve the
companion.  We should therefore consider whether the companion that
we have resolved (and named HD~129333B) is identical to the
spectroscopically inferred one.

\subsubsection{The Combined Radial Velocity and Astrometric Solution
\label{sec_rvastrom}}

Combining r.v.\ and astrometric data presents a powerful
approach to fully constrain all the orbital elements of a binary
system.  In this Section we test the hypothesis that the DM91 and the 
imaged companions are identical by attempting to solve for the
parameters of the relative orbit and checking for consistency with
all available data.

The orbital parameters that can be determined 
through r.v.\ monitoring of an SB1 are the
eccentricity $e$, the period $P$, the epoch $T_0$ of periastron, the
longitude $\omega$ of periastron, the systemic radial
velocity $v_{\rm rad, 0}$, and the primary velocity semi-amplituide $K_1$.
$K_1$ is related to the other orbital parameters through the mass
function
\begin{equation}
f(M) = \frac{(M_2 \sin i)^3}{(M_1+M_2)^2} = 
\frac{(1-e^2)^{3/2} P K_1^3}{2 \pi G} \label{eqn_fm},
\end{equation}
where $M_{1,2}$ are the component masses, $i$ is the inclination 
of the orbit with respect to the observer, and $\pi$ and $G$ are constants
\citep[e.g.,][p.80]{heintz78}.  The orbital inclination $i$
cannot be constrained from r.v.\ monitoring; hence the masses and
the semi-major axes $a_{1,2}$ of the binary components are degenerate 
with $i$.

From astrometric observations we can fit for $e$, $P$, $T_0$, $\omega$,
$i$, $a_{1,2}$ (and hence, $M_{1,2}$), and for the only remaining 
parameter~--- the longitude of the ascending node $\Omega$.  Therefore,
by performing a least-squares fit to the combined and
appropriately-weighted
r.v.\ and astrometric data, one can fully determine the orbit of a 
binary and attain greater precision in estimating the orbital elements 
\citep{morbey75}.

We first list the orbital parameters that have been already
determined.  Based on the r.v.\ measurements shown in
Figure~\ref{fig_rv}, DM91
find $e=0.665\pm0.023$, $T_0$~= JD~2446932$\pm$20~=
year 1987.37, $\omega=188.0\pm5.2\degr$, $K_1=5.09\pm0.20$~km~s$^{-1}$,
and $P=4575$~days~= 12.53~years.  DM91 state however,
that the period is probably accurate only ``to the nearest unit 
of $\log P$'' (i.e., $10^{3.5}<P<10^{4.5}$~days, or between 8.7 and
87 years), and calculate the uncertainties in $e$, $T_0$, $\omega$, and
$K_1$ for a fixed $P$.  Nevertheless, because of the high eccentricity 
of the orbit and because of their adequate observational coverage of 
the star near r.v.\ minimum, the final values of these three 
parameters are not expected to be significantly different.
Assuming that the r.v.\ and the resolved companions are the same, we 
impose the additional constraints derived from our astrometric observations,
namely, the separation $\rho$ and position angle $\phi$ between the
binary components on 2003 May 13 ($T$~= JD~2452772~= year 2003.36).
Given the long (multi-year) orbital
period and the small change (insignificant within the error bars) 
between our two relative astrometric observations taken four months apart, 
we only 
use one of the astrometric measurements.  Finally, we adopt a mass of
1.05~$M_\sun$ for the G0~V primary, based on an estimate from
\citet{dorren_guinan94}.  

The equation that determines the binary orbit is Kepler's equation:
\begin{equation}
E-e \sin E = \frac{2\pi}{P}(T-T_0),	\label{eqn_kepler}
\end{equation}
where the eccentric anomaly $E$ is related to the true anomaly $\theta$
through
\begin{equation}
\tan \frac{\theta}{2} = \left (\frac{1+e}{1-e} \right)^{1/2}
\tan \frac{E}{2}.			\label{eqn_anomalies}
\end{equation}
The remaining equations are:
\begin{eqnarray}
P = 2\pi \sqrt{\frac{a^3}{G(M_1+M_2)}} \\
r = a(1-e \cos E) \\
\rho^2 = r^2 (1-\sin^2(\theta+\omega)\sin^2 i) \label{eqn_rho} \\
\tan \Omega = \frac{\tan\phi - \tan(\theta+\omega)\cos 
i}{1 + \tan\phi\tan(\theta+\omega)\cos i}.
\label{eqn_Omega}
\end{eqnarray}
Because the orbital period of HD~129333 is 
poorly constrained by DM91, we choose to treat $P$ as a free parameter.
Thus, the unknown parameters are eight: $P$, $M_2$, $i$, $a$ (the
semi-major axis of the relative orbit), $\Omega$, $E$,
the radius vector $r(T-T_0)$, and the true anomaly $\theta(T-T_0)$ of the 
companion in the relative orbit at time $T-T_0$.  From the combined
imaging and r.v.\ data we have imposed seven constraints: $\rho(T-T_0)$, 
$\phi(T-T_0)$, $e$, $\omega$, $M_1$, $K_1$, and $T-T_0$.
Given that the number of unknowns is greater than the number of
constraints, we cannot solve unambiguously for the parameters, let alone
use a least-squares approach to determine their
best-fitting values.  However, by stepping through a grid of constant 
values for one of the parameters, we can determine the rest.

We choose $M_2$ as our step parameter for the grid, treating it as a
known parameter.  In principle
we can use Equations~\ref{eqn_anomalies}--\ref{eqn_rho} to express 
$P$ in terms of $M_2$, $E$, and the known variables, and then substitute
this expression in Equation~\ref{eqn_kepler}, which can be solved for $E$.
However, because of the complexity of the general functional form
$P(M_2, E, \rho, \phi, e, \omega, M_1, K_1, T-T_0)$, and because Kepler's
equation cannot be solved analytically, we employ a two-stage iterative
approach.  In the outer iteration, for a given value of $M_2$ we converge
upon a solution for $P$, and in the inner iteration we use the 
Newton-Raphson method to solve Kepler's
equation for $E$.  The iterative Newton-Raphson method has been
described in detail elsewhere \citep[e.g.,][]{press_etal92}, so we
do not discuss it further.  The convergence of the outer iteration loop
however merits a brief description.

We take an initial estimate $P_0$ for the period and invert
Equation~\ref{eqn_fm} to obtain a numerical value for $\sin i$ as a
function of $P$ and $M_2$.
From Equations \ref{eqn_anomalies}--\ref{eqn_rho} we then express $P$ in
terms of the known parameters plus $M_2$, $E$, and $\sin i$, and plug 
that expression in
Equation~\ref{eqn_kepler}, which is then easier to solve for $E$.  Once
$E$ is obtained, we invert Kepler's equation to find a solution $P=P_1$
for the orbital period that depends on the initial guess $P_0$.
We repeat the above procedure by substituting $P_1$ for $P_0$, and so
forth until the values $P_j$ converge.
We stop when the value of $P$ is constrained to better than 0.1\%,
which usually occurs after 3--4 iterations.  Because of
the monotonic dependence of the orbital elements $i$, $a$, and $E$ on 
$P$, we can be certain that the thus-obtained solution for $P$ is unique.

Following the above procedure and adopting the DM91 values for $e$,
$T_0$, and $K_1$, we find that if the resolved companion is identical to
the r.v.\ one, its mass is at least $0.68M_\sun$, with a corresponding
period of 50.0 years, $a=16.3$~AU, and $i=85\degr$.  Values as small 
as $M_2=0.58M_\sun$ ($P=42.8$~years, $a=14.4$~AU, $i=84\degr$) 
are possible if all parameters are set at their
one-sigma deviations that minimize $M_2$.  

However, a minimum mass of $\approx0.58M_\sun$ for HD~129333B does 
not agree with the constraint from our IR spectroscopy,
$M_2\leq0.50M_\sun$ (Sections~\ref{sec_spectroscopy},
\ref{sec_masses}), obtained from comparison to theoretical evolutionary
tracks from \citet{baraffe_etal98}.  
Moreover, a companion with mass $M_2\geq0.58M_\sun$ 
\citep[spectral type K8 or earlier;][]{cox00} would be too bright
in absolute magnitude \citep[$M_K\leq5.1$;][]{bessell91} to be identified
with HD~129333B ($M_{K_S}=6.30\pm0.12$, from its apparent magnitude and
from the {\sl Hipparcos} distance to HD~129333).  It is therefore likely
that the r.v.\ and the spectroscopic companion are not identical.

The inconsistency between the masses could be explained by noting that
a recent study of low-mass binaries by
\citet{hillenbrand_white04} has shown that most modern stellar 
evolutionary models tend to underestimate dynamical masses of main and
pre-main sequence stars.  According to the authors,
the \citet{baraffe_etal98} models
underestimate the mass of a $0.5M_\sun$ main sequence star by $\approx$20\%.
This could reconcile the limits on the mass of HD~129333B obtained from
near IR spectroscopy with those from the orbital solution.  However, the
problem of the companion being sub-luminous remains.

\subsubsection{Comparison to Other Radial Velocity Data}

\begin{figure}
\plotone{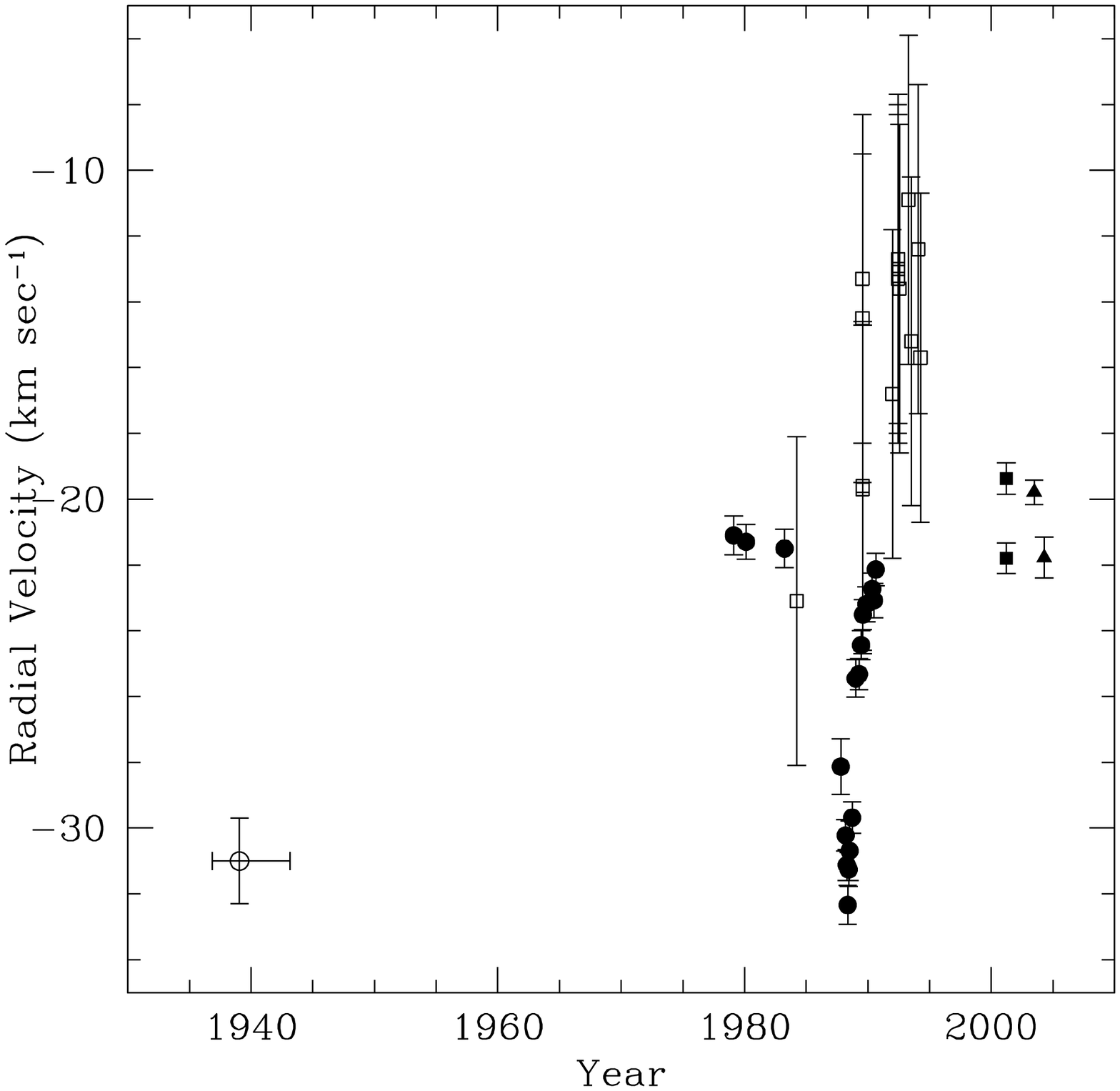}
\figcaption{Radial velocity data for HD~129333 from the literature.
Data from DM91 are plotted as filled circles, from
\citet{dorren_guinan94} as open squares, from \citet{montes_etal01b} as
filled squares, and from \citet{wilson_joy50} as an open circle.  Our
own unpublished r.v.\ data are shown as solid triangles.  The exact epoch of
the \citeauthor{wilson_joy50} observation is unknown.  Given the DM91
orbital solution and the current
phase coverage, periods $<$16~years can be excluded.
\label{fig_rv}}
\end{figure}

The DM91 set of r.v.\ data are the most accurate and deterministic for
the orbit of HD~129333.  Other data exist from \citet{wilson_joy50}, 
\citet{dorren_guinan94}, \citet{montes_etal01b}, and from
our own high-resolution spectroscopic observations\footnote{$v_{\rm rad} 
= -19.79\pm0.37$~km~s$^{-1}$ and $-21.77\pm0.62$~km~s$^{-1}$
on 2002 April 18, and 2003 February 10, respectively (Hillenbrand 
et al., in prep.).} (Figure~\ref{fig_rv}); however, they do not
improve the orbital phase coverage greatly.  Although
\citeauthor{dorren_guinan94} (data plotted as open squares) appear to 
have captured the binary near an r.v.\ maximum around 1993, their data 
are less restrictive because of their large uncertainties.  In addition,
the \citeauthor{dorren_guinan94} data for 1990 systematically 
overestimate the r.v.\ of the 
primary with respect to DM91 data taken over the same period.
We therefore choose to disregard the
\citeauthor{dorren_guinan94} data set.  The remaining data are
very limited and we do not attempt to use them to re-fit the
DM91 orbital solution.  However, they are of some utility in
constraining the orbital period.

Because no other r.v.\ minimum is observed for HD~129333 between 1987.37
and 2003.36, we conclude that $P>16$~years.  The \citeauthor{wilson_joy50} 
data point (based on three measurements) is consistent with an r.v.\ 
minimum, and is thus critical in constraining the orbital period.  
However, the authors do not list an epoch for the observations. 
The \citeauthor{wilson_joy50} data were taken in the course of the Mt.\
Wilson stellar spectroscopic survey, and are kept in the
Ahmanson Foundation Star
Plates Archive\footnote{Maintained at the Carnegie Observatories
of Washington, Pasadena, California.}.  After consultation with the 
original plates, we
retrieve the dates of the individual observations: 1936 March 10 (year
1936.19), 1936 Jun 4 (year 1936.42), and 1942 Jun 24 (year 1942.48).
We adopt the mean date of these observations, the year 
$1938.4_{-2.2}^{+4.1}$, as the epoch for the
\citeauthor{wilson_joy50} measurement $v_{\rm rad}=-31.0\pm1.3$~km~s$^{-1}$,
where the errors in the epoch correspond to the interval between their 
first and 
last observation.  Because of its highly eccentric orbit, the star must
have been within 1 year of r.v.\ minimum at this epoch.  Given the r.v.\ 
minimum in 1987.37 and $P>16$~years, we infer that the r.v.\
companion has a likely orbital period equal to the interval between the
two observed minima, or to some integer fraction thereof: 
$49.0_{-4.2}^{+2.4}$ years, $24.5_{-2.1}^{+1.2}$ years, or 
$16.3_{-1.4}^{+0.8}$ years (all consistent with the DM91 estimate).  

The 49-year orbital period agrees with the one obtained in 
Section~\ref{sec_rvastrom}, and supports the evidence that the r.v.\ and
the resolved companion may be identical.  If the r.v.\ companion was on
a 24.5-year period ($a=9.8$~AU, from the DM91 orbital elements),
it would have been $\gtrsim0.33-0.44\arcsec$ from
the primary during our imaging in 2003, with mass
$M_2\geq0.47M_\sun$.  Such a companion should have been at least as 
bright as the resolved one ($\Delta K_S=3.0$), although could have fallen
just below our detection limits ($\Delta K_S \approx3.0$ at
$0.4\arcsec$) if it was at the lower limit of the allowed mass range.  
Given $q=M_2/M_1\geq0.46$ in this case, the star should be
easily detectable as a double-lined spectroscopic binary (SB2) through
high-resolution spectroscopy in the near IR, where the contrast
favors detecting SB2 systems with mass ratios as small as 0.2
\citep{prato_etal02}.  A 16.3-year period can most probably be
excluded, since the 2003 data point does not indicate an approaching 
r.v.\ minimum (Figure~\ref{fig_rv}) in late-2003--2004, as would 
be expected at this periodicity.

Therefore, even after consideration of additional archival r.v.\ data, the
question about the multiplicity of HD~129333 remains open.  The system
can be either a binary with a 14--16~AU semi-major axis (but a 
discrepancy in the inferred mass of the secondary), or a triple with a 
10~AU inner (spectroscopic) and $\sim$25~AU outer (resolved) components.  
Indeed, SIMBAD does list HD~129333 as a BY~Dra variable, which may 
indicate that the high level of chromospheric activity arises from close 
binarity, rather than extreme youth.  However, the high photospheric 
\ion{Li}{1} abundance of HD~129333 and its kinematic
association with the Pleiades moving group (Section~\ref{sec_masses})
confirm its young age.  Moreover, at 10~AU semi-major axis the 
inferred spectroscopic companion is too distant to be synchronized with
the rotation period of the primary \citep[2.7~days;][]{dorren_guinan94},
and to thus affect its chromospheric activity.  An additional close-in 
component would be required, that should
produce a short-period SB1 or SB2 spectroscopic signature, as in binary 
BY~Dra systems.  Such is not reported by DM91, however.

In deciding whether a triple system with a 24.5-year period for the r.v.\
(inner) companion is a
likely state for HD~129333, it is worth considering the dynamical 
stability of such a system.  We adopt masses of 1.05~$M_\sun$, 
0.5~$M_\sun$ and 0.2~$M_\sun$ for the primary, the
inner, and the outer (resolved) companion, respectively, and
apply a dynamical stability criterion from the numerical analysis of
\citet{donnison_mikilskis95}.  Assuming prograde orbits, the
\citeauthor{donnison_mikilskis95} condition for stability as applied to
HD~129333 states that the distance of closest approach
of the outer companion to the barycenter of the system should be $>$27~AU.  
Variations in the masses of the two companions within the determined 
limits do not change this requirement by more than 3--5~AU.  At a
projected separation of $25.0\pm1.5$~AU from the primary, the 
resolved companion is fully consistent with this requirement.  Hence,
the system can be a dynamically stable triple.  

\subsection{HD 49197B: a Rare Young L Dwarf}

Our empirical knowledge of the photospheres of young ($<$1~Gyr) L dwarfs 
is currently very limited.  The only confirmed such dwarfs are
all companions to main sequence stars: G~196--3B \citep[L2,
20--300~Myr;][]{rebolo_etal98}, Gl~417B\footnote{Gl~417B is itself
considered to be resolved by \citet{bouy_etal03} into two components
with a 70~mas separation, equal to the diffraction limit of their {\sl
HST/WFPC2} observations.} \citep[L4.5, 
80--300~Myr;][]{kirkpatrick_etal01}, the pair HD~130948B/C 
\citep[L2, 300--600~Myr;][]{potter_etal02}, and now HD~49197B
(L4, 260--790~Myr).  It is useful to expand the sample of young L dwarfs
in order to study gravity-sensitive features in their spectra, and to
provide constraints for evolutionary models of ultra-cool dwarfs.

Younger L dwarfs have been reported in several open clusters:
$\sigma$~Ori \citep[1--8~Myr;][]{zapatero_osorio_etal99}, the Trapezium 
\citep[$\sim$1~Myr;][]{lucas_etal01}, and Chameleon~I
\citep[1--3~Myr;][]{lopez_marti_etal04}.  However, these results have not
been independently confirmed.  The youth of 
$\sigma$~Ori~47 (L1.5), and hence its association with the cluster, has
been recently brought into question by \citet{mcgovern_etal04}, who find
that the object shows strong \ion{K}{1} absorption at $J$ band,
characteristic of several Gyr old field dwarfs.
\citet{lucas_etal01} determine M1--L8 spectral types for their objects
in the Trapezium, using water indices defined
for the $R\approx30$ $H$-band spectra.  They also use
\citet{burrows_etal97} sub-stellar evolutionary tracks to infer masses
from $IJH$ photometry.  However, the deduced spectral types and the 
masses correlate very poorly~--- a
result potentially traceable to the anomalous continuum shapes of
their spectra (their Figure~4), some of which appear to contain residual 
telluric or instrument-transmission features (as seen in their
Figure~1) that the authors interpret as
photospheric water absorption.  Finally, in their analysis of
photometrically-identified brown dwarfs toward Chameleon~I,
\citet{lopez_marti_etal04} acknowledge that the
classification of their early L dwarfs is uncertain, because their
locus overlaps with that of extincted M-type objects on
their color-magnitude diagram (Figure~8 in that paper).

Therefore, because of its association with a young star, HD~49197B is 
one of only five known young L dwarfs whose age can be determined with
reasonable certainty.  All five span a narrow range in spectral type: L2--L5.
A program of uniform spectroscopic observations of these 
young L dwarf companions, undertaken in a manner similar to the NIRSPEC
brown dwarf spectroscopic survey of \citet{mclean_etal03},
promises to establish gravity-sensitive standards \citep[as in][for
F8--M7 stars]{kleinmann_hall86} to use in determining
the ages of isolated L dwarfs.

\subsection{Sub-Stellar Companions to Main-Sequence Stars}

Until recently, only a handful of brown dwarf companions 
to nearby A--M stars were known from direct imaging, all at angular
separations $>$4$\arcsec$ \citep[see compilation in][]{reid_etal01}.
With AO technology still in
its early developing stages, ground-based direct imaging observations of 
main-sequence stars were sensitive mostly to massive, widely-separated
sub-stellar companions.  From the observed dearth of brown dwarf
companions to main sequence stars at separations comparable to those in 
main sequence binaries, it was inferred that the
radial-velocity ``brown dwarf desert'' \citep[for separations 
$\lesssim$3~AU;][]{campbell_etal88, marcy_benitz89, marcy_butler00}
extended to at least 120~AU \citep{oppenheimer_etal01},
or 1200~AU \citep{mccarthy01, mccarthy_zuckerman04}, with estimates for the
brown dwarf companion frequency around 1\% within this separation range.  
From 2MASS data however, \citet{gizis_etal01} found that
the brown dwarf 
companion fraction was much higher ($\sim$18\%) at separations
$>$1000~AU from F--M0 dwarfs, and consistent with
that of stellar companions to G stars \citep{duquennoy_mayor91}.  
Though the \citeauthor{gizis_etal01} result is based on only
3 bound companions out of 57 then known field L and T dwarfs \citep*[a
fourth bound companion, GJ~1048B, has now been confirmed in the same 
sample by][]{seifahrt_etal04}, they
exclude a brown dwarf companion fraction of 1.5\% at the 99.5\%
confidence level.
Such an abrupt change in the frequency of bound brown dwarfs at 1000~AU
from main-sequence stars is not predicted by any of the current brown 
dwarf formation scenarios.  More likely 
would be a continuously varying sub-stellar companion fraction
from inside the r.v.\ brown dwarf desert at $<$3~AU out to distances
$>$1000~AU.

Recent results from more
sensitive space- and ground-based surveys 
point to a somewhat higher frequency of sub-stellar companions.
In a survey of 45 young stars within $\sim$50~pc, the NICMOS
Environments of Nearby Stars team has reported the discovery of 2
confirmed brown dwarfs, TWA~5B \citep{webb_etal99, lowrance_etal99} and 
HR~7329B
\citep{lowrance_etal00}, and a probable third one: the binary companion
Gl~577B/C, whose components likely span the stellar/substellar
boundary \citep{lowrance_etal03}.  A similar survey of twenty-four 
5--15~Myr old
stars in the more distant ($\approx$150~pc) Scorpius-Centaurus association
does not detect any plausible sub-stellar companions
\citep{brandner_etal00}.  Even so, the fraction
of stars with sub-stellar companions detected with NICMOS (2--3 out of 69) is 
markedly
higher than the one reported from the two initial large-scale ground-based
surveys \citep[2 out of $\approx$390;][]{oppenheimer_etal01, 
mccarthy_zuckerman04}, 
and is inconsistent with the incompleteness-corrected estimate ($\leq$2\%)
of \citet{mccarthy_zuckerman04}.  Furthermore, as a result of improvements in 
existing AO technology, and the equipping of several large 
telescopes with newly-designed high-order AO systems, recent
ground-based direct imaging efforts have been more successful in
detecting close-in brown dwarf companions to sun-like primaries: 
Gl~86B \citep{els_etal01},  
HD~130948B/C \citep{potter_etal02}, HR~7672B \citep{liu_etal02}, and
HD~49197B (this paper).  All of these are at angular separations 
$<$3$\arcsec$, and at projected distances $<$50~AU from their primaries, 
and hence inaccessible for imaging by \citet{mccarthy_zuckerman04}, whose
survey targeted the 75--1200~AU separation range.  Finally, with the
natural guide star limit of AO systems being pushed to ever fainter 
magnitudes using curvature sensors \citep[down to $\sim$16~mag at 
0.8$\micron$;][]{siegler_etal03}, a number
of very low-mass (VLM) binaries has become known, with separations as small 
as 1~AU.  The components in these VLM binaries
often straddle the stellar-substellar boundary \citep[for a compilation,
see Table~4 in][]{close_etal03}.

The emergent picture from these recent discoveries is that of potential
deficiency of brown dwarfs at 10--1000~AU separations from main sequence
stars, though not as pronounced as in the r.v.\ brown dwarf desert 
\citep[frequency $<$0.5\%;][]{marcy_butler00}.
Based on one detection (of a binary brown dwarf companion) among 
31 stars, \citet{potter_etal02} set a lower
limit of 3.2$\pm$3.2\% for the frequency of brown
dwarfs at 10--100~AU from main sequence stars.
At separations $>$50~AU, from the NICMOS discoveries
\citep{lowrance_etal99, lowrance_etal00} and from their newly-reported
brown dwarf companion to the star GSC~08047--00232 in Horologium,
\citet{neuhauser_guenther04} report that brown
dwarfs are found around 6$\pm$4\% of stars.  The outer scale for the
\citet{neuhauser_guenther04} estimate is not specified, but is probably
limited to 1000--2000~AU 
by the FOV of high angular resolution IR detectors (up to
40$\arcsec$$\times$40$\arcsec$; e.g., NICMOS, or ones used behind AO),
and by the distances out to which young stars are probed for sub-stellar
companions (out to 100--200~pc).  By combining these estimates with the 
\citet{gizis_etal01} estimate of 18$\pm$14\% at separations $>$1000~AU, 
we can conclude that, despite the small number statistics involved, 
there possibly exists a continuum in the frequency distribution of brown dwarf
companions at separations ranging from within the r.v.\ brown dwarf desert
($\leq$3~AU) out to $>$1000~AU (where brown dwarf companions are as
common as stellar ones).  The observed decline in the rate of occurrence
of directly imaged brown dwarf companions at small separations is likely
at least partially an effect of the limited sensitivity of imaging surveys 
to close-in low-mass brown dwarfs.  New, sensitive surveys
for sub-stellar companions, such as the Palomar AO Survey of Young Stars 
are poised to explore this regime in the next few years.

\section{CONCLUSION}

We have presented the observing strategy and first results from the 
Palomar Adaptive Optics
Survey of Young Stars, aimed at detecting sub-stellar companions to
$<$400~Myr solar analogs within 160~pc of the Sun.  We have discovered
low-mass (0.04--0.5~$M_\sun$) companions 
to 4 young nearby stars.  The L4$\pm$1 brown dwarf HD~49197B and the 
M2$\pm$1~V star HD~129333B have been confirmed as companions to their
corresponding primaries through follow-up astrometry
and spectroscopy.  Physical association in the
V522~Per and RX~J0329.1+0118 systems, containing respectively M4$\pm$1 
and M3$\pm$2 secondaries, has 
been established with $>$99.95\% confidence from spectroscopy and
from the expected space density of objects of similar spectral
type.  

The astrometry for the resolved stellar companion to
HD~129333 is found to be consistent with archival r.v.\ data for this
single-lined spectroscopic binary, indicating that the resolved and the
r.v.\ companions may be identical.  Given the inferred mass constraints
on the secondary however, the companion is then underluminous by at least
1~mag at $K_S$.  A solution in which the star is a triple is also
likely.  It does not suffer from similar inconsistencies, and could be
dynamically stable.  Because the expected mass ratio between the inner
two companions of the triple is $\geq$0.46, they should be resolved as a
double-lined spectroscopic binary from high resolution infrared
spectroscopy.  In either case HD~129333 is confirmed to be a multiple
star, and hence not a true analog of the young sun, as previously 
considered \citep[e.g.,][]{dorren_guinan94,strassmeier_rice98}.

The newly-discovered brown dwarf HD~49197B is one of very
few confirmed young ($<$1~Gyr) L dwarfs.  It is also
a member of a small number of brown dwarf companions
imaged at projected separations of $<$50~AU from their host stars,
i.e., at distances comparable to the giant-planet zone in the Solar
System.  The number of such companions, albeit small, has been growing 
steadily in
recent years with the results of more sensitive imaging surveys coming
on-line.  Longer duration radial velocity surveys and
improvements in AO techniques are expected to further push the
detection limits of each method to the point where their
sensitivities overlap.  Although the true extent and depth of the so-called
``brown dwarf desert'' will not be revealed until that time, increased 
sensitivity to sub-stellar
companions at small separation has already resulted in upward revisions
of their estimated frequency.

\begin{acknowledgements}

We would like to thank Richard Dekany and Mitchell Troy for sharing with
us their expertise of the Palomar AO system, Rick Burress and Jeff Hickey
for assistance with PHARO, Randy Campbell, Paola Amico and David
Le Mignant for their guidance in using Keck AO,
Keith Matthews and Dave Thompson for 
help with NIRC2, and our telescope operators Jean Mueller at the Palomar
5~m telescope, and Chuck Sorenson at the Keck~II telescope.  
We are grateful to Tom Hayward, Stephen Eikenberry, and Matthew Britton
for sharing with us their ideas on AO data reduction and PSF
subtraction.  We thank Russel White for insightful discussions on 
spectroscopic binaries and for a critical review of the manuscript, and
the referee for detailed suggestions on chance
alignment probabilities.  We thank
George Carlson and Donna Kirkpatrick of the Ahmanson Foundation Star
Plates Archival Project
for giving us access to the Mt.\ Wilson spectroscopic survey plates.
This publication makes use of data products from the Two Micron All Sky
Survey, which is a joint project of the University of Massachusetts and
the IPAC/California Institute of Technology, funded by the NASA and the NSF.
Use of the FEPS Team database has proven invaluable throughout
the course of our survey.  We thank John Carpenter for building and
maintaining the database.  Finally, the authors wish to extend special
thanks to those of Hawaiian ancestry on whose sacred mountain of Mauna
Kea we are privileged to be guests.  Without their generous hospitality,
some of the observations presented herein would not have been possible.

\end{acknowledgements}
 

\clearpage

\begin{deluxetable}{ccccc}
\tabletypesize{\scriptsize}
\tablewidth{0pt}
\tablecaption{Observations. \label{tab_observations}}
\tablehead{\colhead{Target} & \colhead{First Epoch} & \colhead{Mode} 
	& \colhead{Second Epoch} & \colhead{Mode} \\
	& \colhead{Telescope} & & \colhead{Telescope} & }
\startdata
HD 49197 & 2002 Feb 28 & $JHK_S$ coronagraphic imaging &
	2003 Nov 9--10 & $JHK_S$ coronagraphic imaging, \\
 & Palomar & & Keck II & $K$ spectroscopy \\
HD 129333 & 2003 Jan 12 & $JHK_S$ non-coronagraphic imaging, & 
	2003 May 13 & Br$\gamma$ non-coronagraphic imaging\\
 & Palomar & $J K$ spectroscopy & Palomar & \\
V522 Per & 2003 Sep 20 & $JHK_S$ non-coronagraphic imaging & 
	2003 Nov 10 & $K_S L^\prime$ non-coronagraphic imaging, \\
 & Palomar & & Keck II & $J K$ spectroscopy \\
RX J0329.1+0118 & 2003 Sep 21 & $JHK_S$ coronagraphic imaging & 
	2003 Nov 10 & $K$ spectroscopy \\
 & Palomar & & Keck II & \\
\enddata
\end{deluxetable}

\begin{deluxetable}{lccccccc}
\tabletypesize{\scriptsize}
\tablewidth{0pt}
\tablecaption{Properties of the observed stars. \label{tab_properties}}
\tablehead{\colhead{Object} & \multicolumn{2}{c}{P.M.\ (mas/year)} &
	\colhead{Parallax} & \colhead{Spectral type} &
	\colhead{$K_S$\tablenotemark{a}} & 
	\colhead{Age\tablenotemark{b}} & \colhead{Notes} \\
	& \colhead{$\mu_\alpha \cos\delta$} & \colhead{$\mu_\delta$}
	& \colhead{(mas)} & & \colhead{(mag)} & \colhead{(Myr)} & }
\startdata
HD 49197 & $-35.12\pm1.05$ & $-48.59\pm0.63$ & $22.41\pm0.87$ & 
	F5 V & $6.067\pm0.024$ & $260-790$ & 1,3 \\
HD 129333 & $-138.61\pm0.61$ & $-11.92\pm0.68$ & $29.46\pm0.61$ &
	G0 V & $5.914\pm0.021$ & $10-100$ & 1,4 \\
V522 Per & $17.6\pm3.0$ & $-26.9\pm2.7$ & $5.46\pm0.20$ & 
	\nodata & $9.352\pm0.024$ & $90\pm10$ & 2,5 \\
RX J0329.1+0118 & $5.4\pm1.1$ & $-5.8\pm1.1$ & $\lesssim10$ & 
	G0 IV & $9.916\pm0.019$ & $\approx120$ & 2,6,7
\enddata
\tablenotetext{a}{From the 2MASS Point Source Catalog \citep{cutri_etal03}.}
\tablenotetext{b}{See Section~\ref{sec_masses}.}
\tablenotetext{\ }{{\sl Notes:} 1.\ Proper motion and parallax from 
{\it Hipparcos}
\citep{perryman_etal97}. 2.\ Proper motion from Tycho~2 \citep{hog_etal00}.
3.\ Spectral type from {\it Hipparcos} \citep{perryman_etal97}. 4.\
Spectral type from \citet{buscombe_foster97}. 5.\ Assumed to be at the
mean {\sl Hipparcos} distance \citep{vanleeuwen99} of the $\alpha$~Per
cluster. 6.\ A distance of at least 100~pc can be inferred for this
young \citep[$\sim$100~Myr;][]{frink_etal97} star from its small proper
motion and its location toward the Taurus star forming region. 7.\
Spectral type from \citet{buscombe98}.}
\end{deluxetable}

\begin{deluxetable}{lccccccc}
\tabletypesize{\scriptsize}
\tablewidth{0pt}
\tablecaption{IR magnitudes and colors of the companions.
\label{tab_photometry}}
\tablehead{\colhead{Object} & \colhead{$\Delta J$} & 
	\colhead{$\Delta H$} & \colhead{$\Delta K_S$} &
	\colhead{$J-H$} & \colhead{$H-K_S$} & \colhead{$K_S$} &
	\colhead{$K_S-L^\prime$}}
\startdata
HD 49197B & $9.6\pm1.2$ & $8.52\pm0.12$ & $8.22\pm0.11$ & 
	$1.2\pm1.2$ & $0.33\pm0.20$ & $14.29\pm0.14$ & \nodata\\
HD 49197``C'' & $6.86\pm0.10$ & $6.82\pm0.09$ & $6.68\pm0.10$ &
	$0.27\pm0.14$ & $0.17\pm0.14$ & $12.75\pm0.10$ & \nodata\\
HD 129333B & $3.38\pm0.10$ & $3.13\pm0.09$ & $3.04\pm0.08$ &
	$0.55\pm0.14$ & $0.19\pm0.12$ & $8.95\pm0.08$ & \nodata\\
V522 PerB & $5.69\pm0.09$ & $5.44\pm0.09$ & $5.16\pm0.09$ &
	$0.63\pm0.13$ & $0.38\pm0.13$ & $14.51\pm0.09$ & $0.15\pm0.18$\\
RX J0329.1+0118B & $4.22\pm0.12$ & $3.86\pm0.08$ & $3.65\pm0.08$ &
	$0.59\pm0.15$ & $0.31\pm0.13$ & $12.85\pm0.09$ & \nodata
\enddata
\end{deluxetable}

\begin{deluxetable}{lcccccc}
\tabletypesize{\scriptsize}
\tablewidth{0pt}
\tablecaption{Astrometry of the companions.
\label{tab_astrometry}}
\tablehead{\colhead{Object} & \multicolumn{2}{c}{Epoch 1} & 
	\multicolumn{2}{c}{Epoch 2} & 
	\multicolumn{2}{c}{Epoch 1 (if non-common p.m.)} \\
	& \colhead{offset (arcsec)} & \colhead{P.A.\ (degrees)}
	& \colhead{offset (arcsec)} & \colhead{P.A.\ (degrees)}
	& \colhead{offset (arcsec)} & \colhead{P.A.\ (degrees)}}
\startdata
HD 49197B & 0.9499$\pm$0.0054 & 78.25$\pm0.40$ & 0.9475$\pm$0.0022 &
	77.60$\pm$0.25 & 0.9029$\pm$0.0037 & 81.87$\pm$0.31 \\
HD 49197``C'' & 6.971$\pm$0.030 & 346.13$\pm$0.34 & 7.016$\pm$0.008 &
	346.50$\pm$0.10 & 6.950$\pm$0.009 & 346.10$\pm$0.10 \\
HD 129333B & 0.7343$\pm$0.0032 & 173.19$\pm$0.35 & 0.7363$\pm$0.0032 &
	173.37$\pm$0.35 & 0.7221$\pm$0.0033 & 180.68$\pm$0.37 \\
V552 PerB & 2.0970$\pm$0.0090 & 194.02$\pm$0.34 & 2.0937$\pm$0.0032 &
	193.91$\pm$0.11 & 2.0991$\pm$0.0032 & 193.93$\pm$0.11 \\
RX J0329.1+0118B & 3.75$\pm$0.05 & 303$\pm$5 & 3.781$\pm$0.016 & 
	303.85$\pm$0.34 & 3.714$\pm$0.070 & 303.7$\pm$1.1
\enddata
\end{deluxetable}

\begin{deluxetable}{lcccccc}
\tablewidth{0pt}
\tabletypesize{\scriptsize}
\tablecaption{Spectroscopic measurements for the companions.
\label{tab_eqw}}
\tablehead{ \colhead{Star} & \colhead{EW(\ion{Na}{1})\tablenotemark{a}} &
	\colhead{EW(\ion{Ca}{1})\tablenotemark{a}} & 
	\colhead{EW(CO)\tablenotemark{a}}
	& \colhead{EW(\ion{K}{1})\tablenotemark{b}} & 
	\colhead{H$_2$O index\tablenotemark{b}} &
	\colhead{$T_{\rm eff, Ca}$\tablenotemark{c}} \\
	& \colhead{(\AA)} & \colhead{(\AA)} & \colhead{(\AA)} & 
	\colhead{(\AA)} & & \colhead{(K)} }
\startdata
HD 49197B & $-0.3\pm0.6$ & $0.2\pm0.4$ & $0.82\pm0.05$\tablenotemark{d} 
	& \nodata & \nodata & $<$3000 \\
HD 129333B & $5.17\pm0.59$ & $4.79\pm0.64$ & $7.26\pm0.57$ & $2.18\pm0.11$ &
	\nodata & 3660 \\
V522 PerB & $5.87\pm0.58$ & $3.11\pm0.50$ & $5.67\pm0.38$ & $1.07\pm0.16$ &
	$0.91\pm0.01$ & 3200 \\
RX J0329.1+0118B & $6.76\pm0.60$ & $3.81\pm0.39$ & $8.63\pm0.17$ & \nodata & 
	\nodata & 3300 
\enddata
\tablenotetext{a}{At $K$ band as defined by \citet{ali_etal95}.}
\tablenotetext{b}{At $J$ band as defined by \citet{gorlova_etal03}.}
\tablenotetext{c}{Calculated from empirical relations relating
$T_{\rm eff}$ to EW(\ion{Ca}{1}) \citep[Table~4 in][]{ali_etal95}.
The sign of the linear coefficient in the \ion{Ca}{1} ``cool'' relation 
of \citeauthor{ali_etal95} has been changed from `--' (as erroneously
listed in their paper) to `+' to match the slope of their empirical
data. \citeauthor{ali_etal95} quote an error of $\pm$300~K for this index.}
\tablenotetext{d}{The CO index for HD~49197B is not in 
\AA, but as defined by \citet{mclean_etal03}.}
\end{deluxetable}

\begin{deluxetable}{lcccc}
\tablewidth{0pt}
\tablecaption{Estimated properties of the companions.
\label{tab_estimates}}
\tablehead{\colhead{Objects} & \colhead{Spectral type} & 
	\colhead{Mass} & \colhead{Projected Separation}
	& \colhead{Probability of} \\
	& & \colhead{($M_\sun$)} & \colhead{(AU)} & \colhead{chance
	alignment}}
\startdata
HD 49197B & L4 $\pm$ 1\tablenotemark{a} & $0.060_{-0.020}^{+0.012}$ & 
	43 & $3\times10^{-6}$ \\
HD 129333B & M2 $\pm$ 1\tablenotemark{b} & $0.20_{-0.08}^{+0.30}$ & 
	25 & $3\times 10^{-6}$ \\
V522 PerB & M4 $\pm$ 1\tablenotemark{b} & $0.125\pm0.025$ & 
	400 & $2\times10^{-3}$ \\
RX J0329.1+0118B & M3 $\pm$ 2\tablenotemark{b} & $0.20_{-0.10}^{+0.30}$ & 
	380 & $9\times10^{-4}$ 
\enddata
\tablenotetext{a}{Inferred from the $K$-band
spectrum, and from the absolute magnitude of the object
(Section~\ref{sec_spec_hd49197b}).}
\tablenotetext{b}{Based on the estimate of the \ion{Ca}{1}-derived 
effective 
temperature (Table~\ref{tab_eqw}), and on the $J$-band \ion{K}{1} and
H$_2$O absorption (if available).  A MK spectral type vs.\ 
$T_{\rm eff}$ classification for dwarfs is adopted from \citet{bessell91}.}
\end{deluxetable}

\clearpage

\end{document}